%% file: main.tex
\documentclass[apjl,ip,twocolumn]{aastex61}
\usepackage{comment}
\usepackage{ifthen}
\usepackage[switch]{lineno}
\modulolinenumbers[5]

\input{newcommand_gen.tex}

\newcommand{\loopand}{\ifnum\value{planetcounter}=2 and \else\fi}
\newcommand{\loopcomma}{\ifnum\value{planetcounter}<2 ,\else. \fi}
\newcommand{\loopcommanoperiod}{\ifnum\value{planetcounter}<2 ,\else \space\fi}
\newcommand{\loopcommanospace}{\ifnum\value{planetcounter}<2 ,\else \fi}

\newcommand{\hatcurrotper}{\ensuremath{41.72 \pm 0.14}}
\newcommand{\hatcurrotpershort}{\ensuremath{41.72}}

\input{HATS755-002.tex}
\input{HATS755-002.hps.tex}

\newcommand{\hatcur}{HATS-71}
\newcommand{\hatcurb}{HATS-71b}

\newcommand{\hatcurLBirhpsmodel}{0.4738}
\newcommand{\hatcurLBiirhpsmodel}{0.2970}
\newcommand{\hatcurLBiRhpsmodel}{0.4237}
\newcommand{\hatcurLBiiRhpsmodel}{0.3158}
\newcommand{\hatcurLBiihpsmodel}{0.2798}
\newcommand{\hatcurLBiiihpsmodel}{0.3962}
\newcommand{\hatcurLBiIhpsmodel}{0.2513}
\newcommand{\hatcurLBiiIhpsmodel}{0.4265}

\newcommand{\hatcurRVgammaabs}{\ensuremath{24.1\pm1.4}}                           

\newcommand{\hatcurCCtwomassshort}{01021226-6145216}

\newcommand{\hatcurCCgaiadrtwoshort}{4710594412266148352}

\newcommand{\hatcurCCgaiamGshort}{15.35}

\newcommand{\hatcurSMEversion}{i}                                       

\newcommand{\hatcurTICID}{TIC~234523599}
\newcommand{\hatcurTOIID}{TOI~127.01}

\newcommand{\hatcurSMEteff}{\ifthenelse{\equal{\hatcurSMEversion}{i}}{\hatcurSMEiteff{}}{\hatcurSMEiiteff{}}}
\newcommand{\hatcurSMEzfeh}{\ifthenelse{\equal{\hatcurSMEversion}{i}}{\hatcurSMEizfeh{}}{\hatcurSMEiizfeh{}}}
\newcommand{\hatcurSMEzfehshort}{\ifthenelse{\equal{\hatcurSMEversion}{i}}{\hatcurSMEizfehshort{}}{\hatcurSMEiizfehshort{}}}
\newcommand{\hatcurSMElogg}{\ifthenelse{\equal{\hatcurSMEversion}{i}}{\hatcurSMEilogg{}}{\hatcurSMEiilogg{}}}
\newcommand{\hatcurSMEvsin}{\ifthenelse{\equal{\hatcurSMEversion}{i}}{\hatcurSMEivsin{}}{\hatcurSMEiivsin{}}}
\newcommand{\hatcurSMEvmac}{\ifthenelse{\equal{\hatcurSMEversion}{i}}{\hatcurSMEivmac{}}{\hatcurSMEiivmac{}}}
\newcommand{\hatcurSMEvmic}{\ifthenelse{\equal{\hatcurSMEversion}{i}}{\hatcurSMEivmic{}}{\hatcurSMEiivmic{}}}

\newcounter{planetcounter}

\newboolean{emulateapj}
\setboolean{emulateapj}{true}

\newboolean{rvtablelong}
\setboolean{rvtablelong}{true}

\newboolean{astroph}
\setboolean{astroph}{true}

\shortauthors{Bakos et al.}
\shorttitle{\hatcur{}\lowercase{b}}
\ifthenelse{\boolean{emulateapj}}{
    
}{
    
}

\begin{document}

\title{
HATS-71b: A giant planet transiting an M3 dwarf star in
TESS Sector 1\footnote{The HATSouth network is operated by a
collaboration consisting of Princeton University (PU), the Max Planck
Institute f\"ur Astronomie (MPIA), the Australian National University
(ANU), and the Pontificia Universidad Cat\'olica de Chile (PUC).  The
station at Las Campanas Observatory (LCO) of the Carnegie Institute is
operated by PU in conjunction with PUC, the station at the High Energy
Spectroscopic Survey (H.E.S.S.) site is operated in conjunction with
MPIA, and the station at Siding Spring Observatory (SSO) is operated
jointly with ANU\@.  This paper includes data gathered with the 6.5 meter
Magellan Telescopes at Las Campanas Observatory, Chile.}
}

\correspondingauthor{G\'asp\'ar Bakos}
\email{gbakos@astro.princeton.edu}

\author[0000-0001-7204-6727]{G.~\'A.~Bakos}
\altaffiliation{Packard Fellow}
\affil{Department of Astrophysical Sciences, Princeton University, NJ 08544, USA}
\affil{MTA Distinguished Guest Fellow, Konkoly Observatory, Hungary}

\author[0000-0001-6023-1335]{D.~Bayliss}
\affil{Department of Physics, University of Warwick, Coventry CV4 7AL, UK}

\author[0000-0002-9832-9271]{J.~Bento}
\affil{Research School of Astronomy and Astrophysics, Australian
	National University, Canberra, ACT 2611, Australia}

\author[0000-0002-0628-0088]{W.~Bhatti}
\affil{Department of Astrophysical Sciences, Princeton University, NJ 08544, USA}

\author[0000-0002-9158-7315]{R.~Brahm}
\affil{Center of Astro-Engineering UC, Pontificia Universidad
	Cat\'olica de Chile, Av.~Vicu\~{n}a Mackenna 4860, 7820436 Macul,
	Santiago, Chile}
\affil{Instituto de Astrof{\'{i}}sica, Pontificia Universidad
	Cat{\'{o}}lica de Chile, Av.~Vicu{\~{n}}a Mackenna 4860, 7820436
	Macul, Santiago, Chile}
\affil{Millenium Institute of Astrophysics, Av.~Vicu{\~{n}}a Mackenna
	4860, 7820436 Macul, Santiago, Chile}

\author{Z.~Csubry}
\affil{Department of Astrophysical Sciences, Princeton University, NJ 08544, USA}

\author[0000-0001-9513-1449]{N.~Espinoza}
\affil{Max Planck Institute for Astronomy, K{\"{o}}nigstuhl 17, 69117 -
	Heidelberg, Germany}
\affil{Bernoulli Fellow}
\affil{IAU-Gruber Fellow}

\author[0000-0001-8732-6166]{J.~D.~Hartman}
\affil{Department of Astrophysical Sciences, Princeton University, NJ 08544, USA}

\author{Th.~Henning}
\affil{Max Planck Institute for Astronomy, K{\"{o}}nigstuhl 17, 69117 -
	Heidelberg, Germany}

\author[0000-0002-5389-3944]{A.~Jord\'an}
\affil{Millenium Institute of Astrophysics, Av.~Vicu{\~{n}}a Mackenna
	4860, 7820436 Macul, Santiago, Chile}
\affil{Instituto de Astrof{\'{i}}sica, Pontificia Universidad
	Cat{\'{o}}lica de Chile, Av.~Vicu{\~{n}}a Mackenna 4860, 7820436 Macul,
	Santiago, Chile}

\author[0000-0002-9428-8732]{L.~Mancini}
\affil{Department of Physics, University of Rome Tor Vergata, Via della
	Ricerca Scientifica 1, I-00133 - Roma, Italy}
\affil{Max Planck Institute for Astronomy, K{\"{o}}nigstuhl 17, 69117 -
	Heidelberg, Germany}
\affil{INAF - Astrophysical Observatory of Turin, Via Osservatorio 20,
	I-10025 - Pino Torinese, Italy}

\author[0000-0003-4464-1371]{K.~Penev}
\affil{Department of Physics, University of Texas at Dallas,
	Richardson, TX 75080, USA}

\author[0000-0003-2935-7196]{M.~Rabus}
\affil{Instituto de Astrof{\'{i}}sica, Pontificia Universidad
	Cat{\'{o}}lica de Chile, Av.~Vicu{\~{n}}a Mackenna 4860, 7820436 Macul,
	Santiago, Chile}
\affil{Max Planck Institute for Astronomy, K{\"{o}}nigstuhl 17, 69117 -
	Heidelberg, Germany}
\affil{Visiting astronomer, Cerro Tololo Inter-American Observatory,
	National Optical Astronomy Observatory.}

\author[0000-0001-8128-3126]{P.~Sarkis}
\affil{Max Planck Institute for Astronomy, K{\"{o}}nigstuhl 17, 69117 -
	Heidelberg, Germany}

\author[0000-0001-7070-3842]{V.~Suc}
\affil{Instituto de Astrof{\'{i}}sica, Pontificia Universidad
	Cat{\'{o}}lica de Chile, Av.~Vicu{\~{n}}a Mackenna 4860, 7820436
	Macul, Santiago, Chile}

\author[0000-0002-0455-9384]{M.~de Val-Borro}
\affil{Astrochemistry Laboratory, Goddard Space Flight Center, NASA,
	8800 Greenbelt Rd, Greenbelt, MD 20771, USA}

\author[0000-0002-4891-3517]{G.~Zhou}
\affil{Harvard-Smithsonian Center for Astrophysics, 60 Garden St.,
	Cambridge, MA 02138, USA}

\author[0000-0003-1305-3761]{R.~P.~Butler}
\affil{Department of Terrestrial Magnetism, Carnegie Institution for
	Science, Washington, DC 20015, USA}

\author{J.~Crane}
\affil{The Observatories of the Carnegie Institution for Science, 813
	Santa Barbara St, Pasadena, CA 91101, USA}

\author[0000-0002-3663-3251]{S.~Durkan}
\affil{Astrophysics Research Centre, Queens University, Belfast,
	Northern Ireland, UK}

\author{S.~Shectman}
\affil{The Observatories of the Carnegie Institution for Science, 813 Santa
	Barbara St, Pasadena, CA 91101, USA}

\author{J.~Kim}
\affil{Department of Astrophysical Sciences, Princeton University, NJ
	08544, USA}

\author{J.~L\'az\'ar}
\affil{Hungarian Astronomical Association, 1451 Budapest, Hungary}

\author{I.~Papp}
\affil{Hungarian Astronomical Association, 1451 Budapest, Hungary}

\author{P.~S\'ari}
\affil{Hungarian Astronomical Association, 1451 Budapest, Hungary}

\author{G.~Ricker}
\affil{Massachusetts Institute of Technology}

\author{R.~Vanderspek}
\affil{Massachusetts Institute of Technology}

\author{D.~W.~Latham}
\affil{Harvard-Smithsonian Center for Astrophysics, 60 Garden St.,
	Cambridge, MA 02138, USA}

\author{S.~Seager}
\affil{Massachusetts Institute of Technology}

\author{J.~N.~Winn}
\affil{Department of Astrophysical Sciences, Princeton University, NJ 08544, USA}

\author{J.~Jenkins}
\affil{SETI Institute/NASA Ames Research Center}

\author{A.~D.~Chacon}
\affil{Millenium Engineering and Integration Co./NASA Ames Research
	Center}

\author{G.~F\H{u}r\'esz}
\affil{Massachusetts Institute of Technology}

\author{B.~Goeke}
\affil{Massachusetts Institute of Technology}

\author{J.~Li}
\affil{SETI Institute/NASA Ames Research Center}

\author{S.~Quinn}
\affil{Harvard-Smithsonian Center for Astrophysics, 60 Garden St.,
	Cambridge, MA 02138, USA}

\author{E.~V.~Quintana}
\affil{NASA Goddard Space Flight Center}

\author{P.~Tenenbaum}
\affil{SETI Institute/NASA Ames Research Center}

\author{J.~Teske}
\affil{The Observatories of the Carnegie Institution for Science, 813
	Santa Barbara St, Pasadena, CA 91101, USA}

\author{M.~Vezie}
\affil{Massachusetts Institute of Technology}

\author{L.~Yu}
\affil{Massachusetts Institute of Technology}

\author{C.~Stockdale}
\affil{Hazelwood Observatory, Victoria, Australia}

\author{P.~Evans}
\affil{El Sauce Observatory, Chile}

\author{H.~M.~Relles}
\affil{Harvard-Smithsonian Center for Astrophysics, 60 Garden St.,
	Cambridge, MA 02138, USA}


\begin{abstract}

\setcounter{footnote}{10}
We report the discovery of \hatcurb{}, a transiting gas giant planet on
a $P=\hatcurLCPshort{}$ day orbit around a $G =
\hatcurCCgaiamGshort$\,mag M3 dwarf star. \hatcur{} is the coolest M
dwarf star known to host a hot Jupiter. The loss of light during
transits is 4.7\%, more than any other 
confirmed transiting planet system. The planet was
identified as a candidate by the ground-based HATSouth transit
survey. It was confirmed using ground-based photometry, spectroscopy, and
imaging, as well as space-based photometry from the NASA {\em TESS}
mission (TIC 234523599). Combining all of these data, and utilizing
Gaia~DR2, we find
that the planet has a radius of $\hatcurPPrlong{}$\,\rjup\ and mass of
$\hatcurPPmlong{}$\,\mjup\ (95\% confidence upper limit of
0.81\,\mjup), while the star has a mass of $\hatcurISOmlong{}$\,\msun\
and a radius of $\hatcurISOrlong{}$\,\rsun. The Gaia DR2 data show
that \hatcur{} lies near the binary main sequence in the Hertzsprung-Russell
diagram, suggesting that there may be an unresolved stellar binary
companion. All of the available data is well fitted by a model in
which there is a secondary star of mass 0.24~$M_\odot$, although
we caution that at present there is no direct
spectroscopic or imaging evidence for such a companion. Even
if there does exist such a stellar companion, the radius and mass
of the planet would be only marginally different from the values we have
calculated under the assumption that the star is single.

\setcounter{footnote}{0}
\end{abstract}



\section{Introduction}
\label{sec:introduction}

Much has been learned about the physical properties of exoplanets in
the nearly three decades following the discovery of the
exoplanet candidate HD~114762\,b \citep{latham:1989}. As of 2018
September 27, the NASA Exoplanet Archive lists 3791 confirmed and validated
exoplanets, the majority of which were found by the NASA {\em Kepler}
mission via the transit method. Among the confirmed planets are
418 short-period gas giant planets ($P < 10$\,days, and $M_{p} >
0.2$\,\mjup\ or $R_{P} > 0.7$\,\rjup). These are the so-called
hot-Jupiters. Especially important are the 375 hot Jupiters which are
known to transit their host stars. These objects are among the best-studied
planets, providing a wealth of information about their physical
properties. Among the 270 planets for which the mass and radius
have both been determined with a precision of 20\% or better, 235 are
hot Jupiters. Of the 133 planets for which the (sky projected) stellar
obliquity has been measured, 117 are hot Jupiters
\citep[TEPCat; ][]{southworth:2011}. Similarly, the
majority of exoplanets with observational constraints on the properties
of their atmospheres are hot Jupiters
\citep[e.g.,][]{madhusudhan:2018}. All of these observations have been
greatly facilitated by the frequently occurring and deep ($\sim1\%$)
transits presented by these systems.

All but twelve of the 418 hot Jupiters in the NASA Exoplanet Archive
have been found around F, G or K-type host stars ($4000\,{\rm K} <
T_{\rm eff} < 7300\,{\rm K}$, or $0.6\,M_{\odot} < M < 1.6\,M_{\odot}$
if $T_{\rm eff}$ is not given in the database).  One of the hot
Jupiters in this sample is around a B-star, seven are around A stars,
and only four have been found around M dwarf stars.  The hot Jupiters
that have previously been discovered around M dwarf stars include
Kepler-45\,b \citep[$M_{P} = 0.505 \pm 0.090$\,\mjup,
$M_{S} = 0.59 \pm 0.06$\,\msun, $T_{\rm eff} = 3820 \pm
90$\,K][]{johnson:2012}, 
HATS-6\,b \citep[$M_{P} = 0.319 \pm
0.070$\,\mjup, $M_{S} = 0.574^{+0.020}_{-0.027}$\,\msun, 
$T_{\rm eff} = 3724 \pm 18$\,K][]{hartman:2015:hats6}, 
NGTS-1\,b \citep[$M_{P} = 0.812^{+0.066}_{-0.075}$\,\mjup, 
$M_{S} = 0.617^{+0.023}_{-0.062}$\,\msun, 
$T_{\rm eff} = 3916^{+71}_{-63}$\,K][]{bayliss:2018:ngts1}, 
and HD~41004\,B\,b \citep[$M_{P}\sin i = 18.37\pm0.22$\,\mjup, $M_{S}
\sim 0.4$\,\msun][]{zucker:2003}. The latter object was detected in
the radial velocity (RV) observations of the M2V component of a K1V+M2V
visual binary, and the inferred 19\,\mjup\ brown-dwarf companion mass
is a lower limit. The other three objects are transiting systems.

Theoretical models of planet formation and evolution have predicted
that hot Jupiters should be less common around M dwarf stars than
around solar-type stars \citep{mordasini:2012}.
While there is some observational support for this prediction from RV
surveys \citep{johnson:2010}, the number of M dwarfs that have been
systematically surveyed for hot Jupiters is still too low to be certain
of this conclusion \citep{obermeier:2016}.

One of the main goals in current exoplanet research is to expand the
sample of well-characterized hot Jupiters known around M dwarfs and A
or earlier-type stars. This will allow the occurrence rate of hot
Jupiters to be measured as a function of stellar mass, and will also
enable the dependence of other planetary system properties on stellar
mass to be studied. Some of these properties that might be
investigated include the orbital obliquities of the planets, the degree
of inflation in the planetary radii, and the atmospheric properties of
the planets.

Giant planets transiting M dwarf stars also provide at least two
observational advantages over similar-size planets transiting larger
stars. They produce very deep transits. In principle, a giant planet
could completely obscure a very low-mass star, although no such system has
been discovered to date. The deep transits allow for observations with
a higher signal-to-noise ratio (S/N), especially if conducted in the IR where
the stars have a higher photon flux density. The stars themselves
undergo very little evolution over the lifetime of
the Galaxy, enabling a more precise constraint on the mass and radius
of the star (and hence of the planet) from the available observations
compared to what can be done for more massive stars 
\citep[e.g.,][]{hartman:2015:hats6}.

The primary challenge in discovering transiting hot Jupiters around M
dwarfs is the faintness of these stars. In order to survey a
sufficient number of M dwarfs to detect the rare cases of
transiting hot Jupiters, it is necessary to observe stars
down to $V \sim 15$\,mag, which is fainter than the limits of many of
the ground-based transit surveys that have been productive at
discovering transiting hot Jupiters. The two ground-based surveys
which have discovered transiting hot Jupiters around M dwarfs are the
HATSouth survey \citep{bakos:2013:hatsouth} and the NGTS survey
\citep{wheatley:2018}. Both of these projects use larger aperture
telescopes compared to the other wide-field transit surveys (0.18\,m in
the case of HATSouth and 0.20\,m in the case of NGTS) allowing for
greater sensitivity to M dwarf stars.

In this paper we present the discovery of \hatcurb{} by the HATSouth
survey, the fifth hot Jupiter found around an M dwarf star, and the
fourth transiting system of this type. With a spectroscopic effective
temperature of \hatcurSMEteff\,K, and a spectral type of M3V, \hatcur{}
is the coolest M dwarf known to host a transiting hot
Jupiter. The 4.7\% deep transits are also the deepest of any
transiting system discovered to date. The planet was first detected by
HATSouth, and then confirmed using ground-based spectroscopic and
photometric follow-up. It was also recently observed in Sector 1 of
the NASA {\em Transiting Exoplanet Survey Satellite} mission ({\em
TESS}, \citealp{ricker:2015}), and included in the first set of alerts
released to the public. In this paper we present all of these data and
analyze them to determine the physical properties of the planet
\hatcurb{} and its host star \hatcur{}. We also present evidence,
driven largely by observations from the Gaia~DR2 mission
\citep{gaiamission,gaiadr2}, that the planet host may have an
unresolved binary star companion with a current projected physical
separation of less than $14$\,AU\@. If confirmed, the presence of this
companion might be responsible for shrinking the orbit of
the gas giant planet to its
current short period orbit.

In Section~\ref{sec:obs} we present the observations. We describe the
analyses that we have performed to confirm the planetary system and
determine its properties in Section~\ref{sec:analysis}. We conclude
with a discussion of the results in Section~\ref{sec:discussion}.

\section{Observations}
\label{sec:obs}

\subsection{Photometric detection}
\label{sec:detection}

\hatcur\ was initially detected as a transiting planet candidate based
on observations by the HATSouth network. A total of 26,668
observations were gathered at 4\,min cadence between UT 2011 July 17
and UT 2012 October 25. The source was observed by the HS-1, HS-3 and
HS-5 instruments (located in Chile, Namibia, and Australia,
respectively) in HATSouth field G755, and by the HS-2, HS-4 and HS-6
instruments (located in Chile, Namibia, and Australia, respectively) in
HATSouth field G756. Observations were carried out as described by
\citet{bakos:2013:hatsouth}, and reduced to trend-filtered light curves
\citep[filtered using the method of][]{kovacs:2005:TFA} and searched
for transiting planet signals \citep[using the Box-fitting Least
Squares or BLS method;][]{kovacs:2002:BLS} as described by
\citet{penev:2013:hats1}. We identified a periodic box-shaped transit
signal in the trend-filtered light curve of \hatcur{} with a period of
$\hatcurLCPshort$\,days\ and a depth of \hatcurLCdip{}\,mmag. Based on
this we selected the object as a candidate, assigning it the HATSouth
candidate identifier \hatcurhtr. The trend-filtered HATSouth light
curve has a residual RMS of 50\,mmag. The light curve is shown
phase-folded in \reffigl{hatsouth}, while the data are made available
in \reftabl{phfu}.

We searched for additional periodic signals in the combined HATSouth
light curve using both the Generalized Lomb-Scargle periodogram
\citep{zechmeister:2009} and the BLS algorithm, in both cases applied
to the light curve after subtracting the best-fit transit model for
\hatcurb{}. We find a peak in the GLS periodogram at a period of
\hatcurrotper\,days with a false alarm probability of $10^{-31}$
(\reffigl{gls}). This false alarm probability is estimated using the
relations from \citep{zechmeister:2009} appropriate for Gaussian
white-noise, but calibrated to the observed sampling and magnitude
distribution via bootstrap simulations. The signal is independently
detected in the G755 and G756 HATSouth light curves (with peak periods
of $37.02$\,days and $41.86$\,days, and false alarm probabilities of
$10^{-10}$ and $10^{-15}$, respectively), which have similar
time-coverage but were obtained with different instruments using
different pointings on the sky. Fitting a sinusoid to the phase-folded
data yields a semi-amplitude of $0.0134 \pm 0.0039$\,mag. We interpret
this period as the photometric rotation period of the star. 
Given the measured rotation period and stellar radius, the
spectroscopic \vsini\ should be $<0.625\,\ms$, i.e., 
undetectable even with the current high-resolution spectroscopy.
Both the period and amplitude are typical values for a field M3 dwarf
star. No additional significant transit signals are detected by BLS in the
combined HATSouth light curve. The highest peak in the spectrum has a
period of $82.7$\,days, a transit depth of $8.5$\,mmag and a
signal-to-pink-noise of only $4.5$.

\ifthenelse{\boolean{emulateapj}}{
    \begin{figure}[!ht]
}{
    \begin{figure}[!ht]
}
\plotone{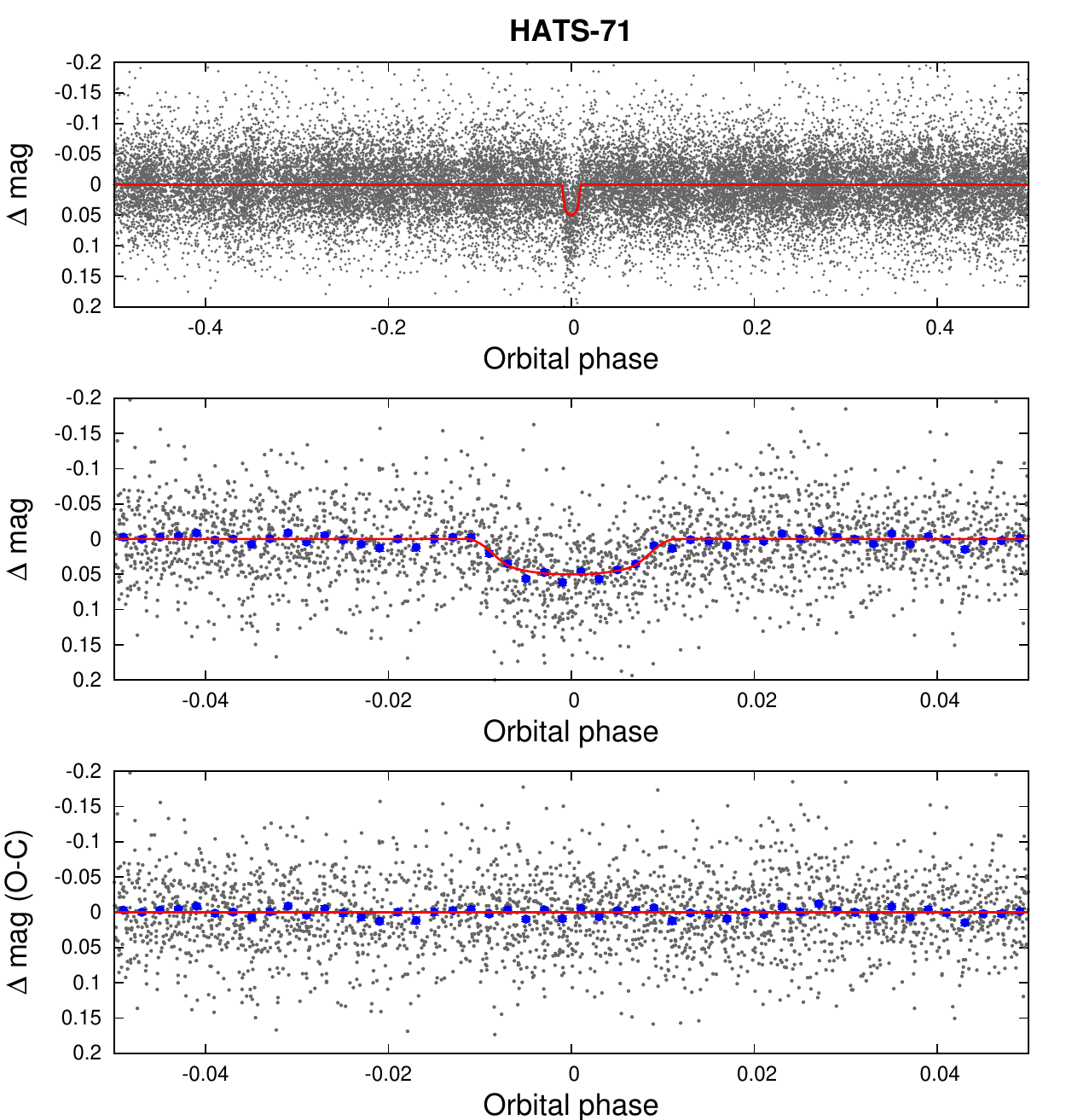}
\caption{
    Phase-folded unbinned HATSouth light curve for \hatcur{}.  {\em
    Top:} the full light curve.  {\em Middle:} the light curve
    zoomed-in on the transit.  {\em Bottom:} the residuals from the
    best-fit model zoomed-in on the transit.  The solid line shows the
    model fit to the light curve.  The dark filled circles show the
    light curve binned in phase with a bin size of 0.002.
\label{fig:hatsouth}}
\ifthenelse{\boolean{emulateapj}}{
    \end{figure}
}{
    \end{figure}
}

\ifthenelse{\boolean{emulateapj}}{
    \begin{figure}[!ht]
}{
    \begin{figure}[!ht]
}
\plotone{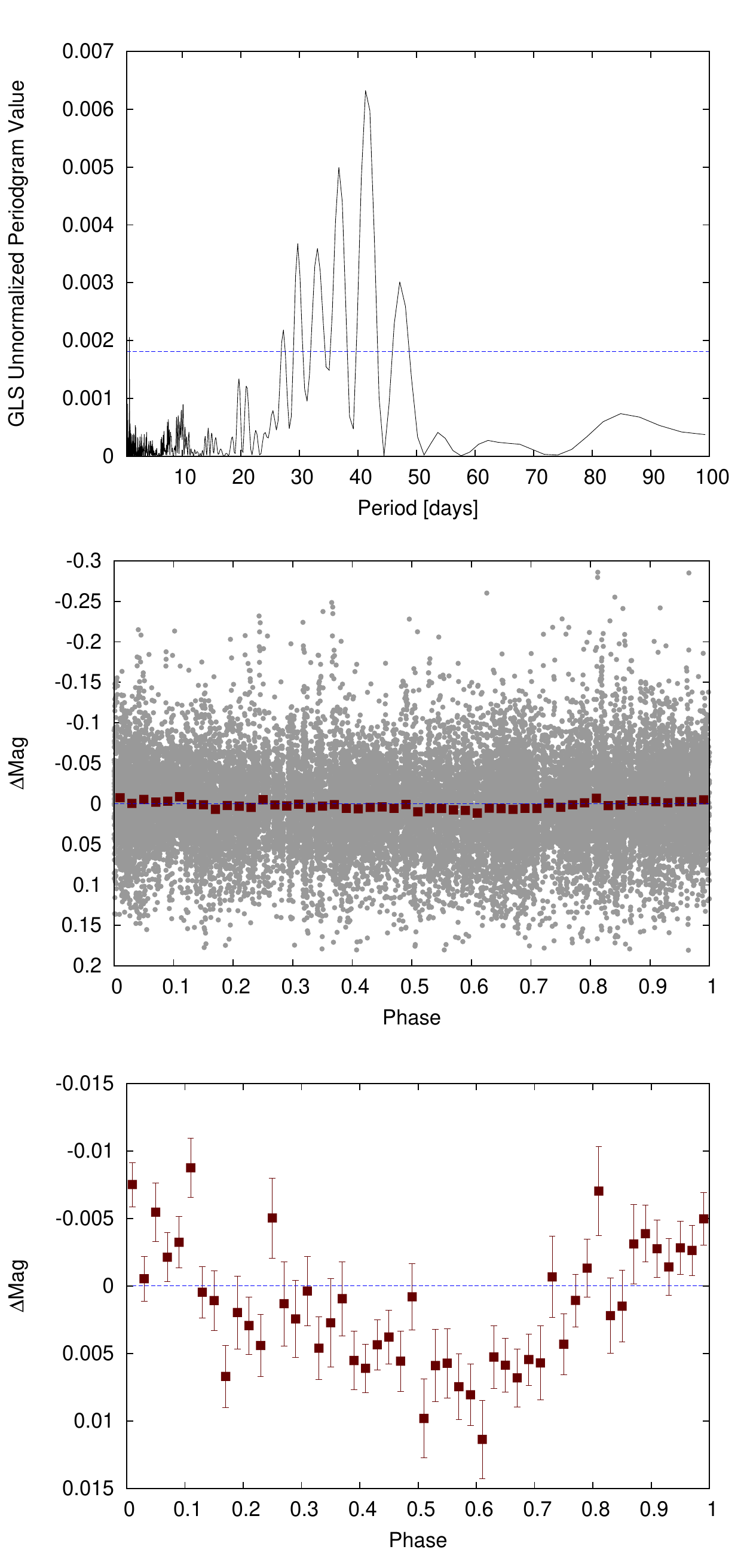}
\caption{
    {\em Top:} Generalized Lomb-Scargle (GLS) periodogram of the
    combined HATSouth light curve after subtracting the best-fit
    transit model for \hatcurb{}. The horizontal dashed blue line shows
    the $10^{-5}$ false alarm probability level. {\em Middle:} The
    HATSouth light curve phase-folded at the peak GLS period of
    \hatcurrotpershort\,days. The gray points show the individual
    photometric measurements, while the dark red filled squares show
    the observations binned in phase with a bin size of 0.02. {\em
    Bottom:} Same as the middle, here we restrict the vertical range of
    the plot to better show the variation seen in the phase-binned
    measurements. 
\label{fig:gls}}
\ifthenelse{\boolean{emulateapj}}{
    \end{figure}
}{
    \end{figure}
}

\subsection{Spectroscopic Observations}
\label{sec:obsspec}

Spectroscopic follow-up observations of \hatcur{} were obtained with
WiFeS on the ANU~2.3\,m \citep[][]{dopita:2007}, PFS on the
Magellan~6.5\,m \citep[][]{crane:2006,crane:2008,crane:2010}, and
ARCoIRIS on the Blanco~4\,m telescope \citep{abbott:2016}. The target
was also observed with FEROS on the MPG~2.2\,m \citep[][]{kaufer:1998}
between 2016 July 1 and 2016 September 16, but the spectra were all too
low S/N to be of use.

%
\ifthenelse{\boolean{emulateapj}}{
    \begin{figure*} [ht]
}{
    \begin{figure}[ht]
}
\plotone{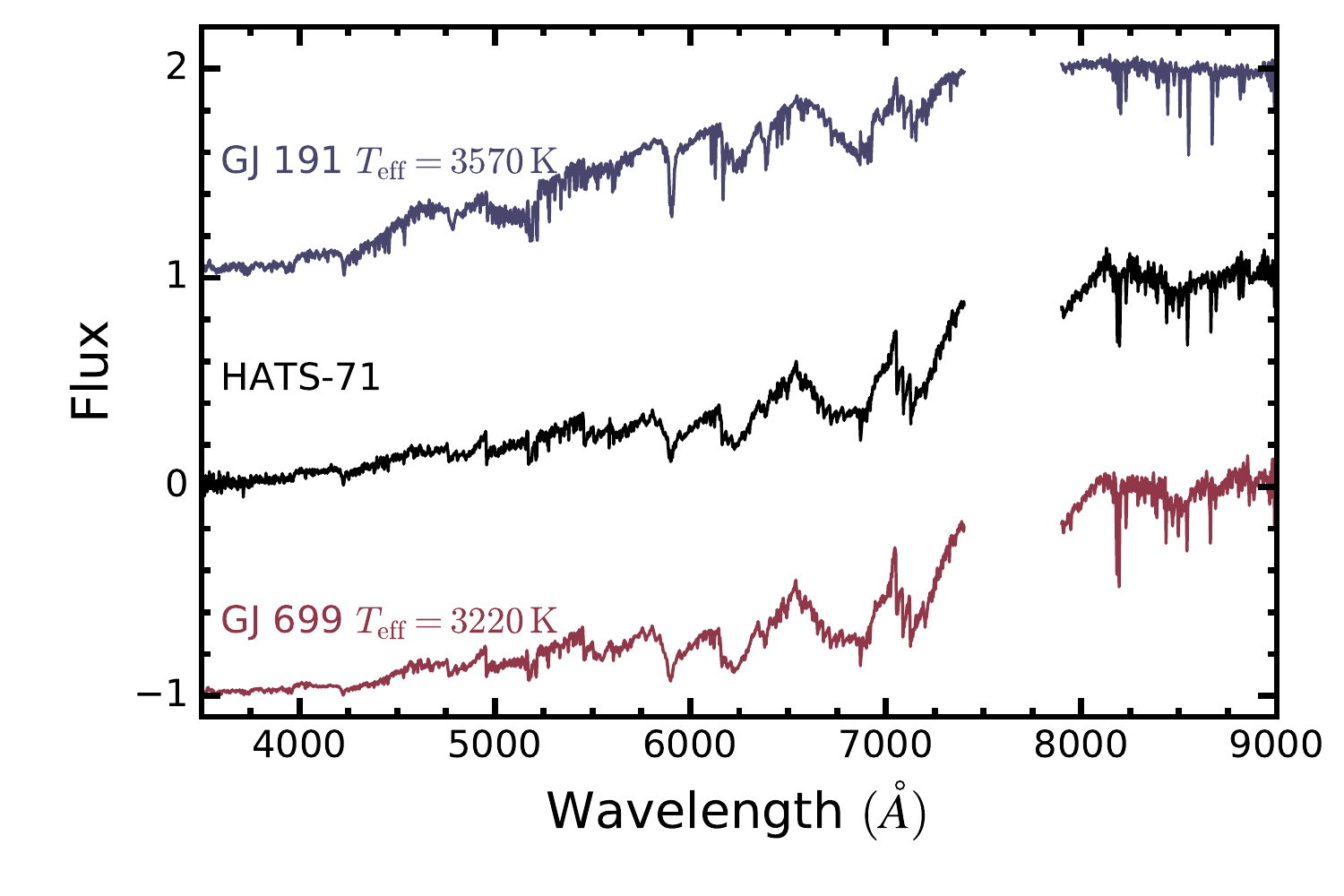}
\caption{
 WiFeS/ANU~2.3\,m $R = 3000$ optical spectra of \hatcur{} (middle
 spectrum) and two other M dwarf standard stars for comparison. 
 \hatcur{} has the optical spectrum of an M3 dwarf star. The
 relative fluxes are on an arbitrary scale, and the two standard
 stars have been shifted vertically for clarity.
}
\label{fig:wifes}
\ifthenelse{\boolean{emulateapj}}{
    \end{figure*}
}{
    \end{figure}
}

The WiFeS observations of \hatcur{}, which were reduced following
\citet{bayliss:2013:hats3}, were used for reconnaissance of this faint
M dwarf. We obtained a single spectrum at resolution $R \equiv
\Delta\,\lambda\,/\,\lambda \approx 3000$ and S/N per resolution
element of 18.9 on UT 2014 August 6 (\reffigl{wifes}). We used this
observation to estimate the atmospheric parameters of the star. The
classification pipeline described by \citet{bayliss:2013:hats3} yielded
parameters of \teffstar$ = 3500 \pm 300$\,K, \logg$ = 4.7 \pm 0.3$
(cgs), and \feh$ = 0.0 \pm 0.5$\,dex, however a comparison to M dwarf
standards indicates a somewhat lower temperature (\reffigl{wifes}). 
Based on spectral matching to BT-Settl models \citep{allard:2011} we
estimate a temperature of 3350\,K. The spectrum reveals this object to
be a single-lined mid-M dwarf star with $\vsini < 50$\,\kms. We also
obtained four spectra at a resolution of $R \approx 7000$ between 2014
August 6--9 which we used to check for any large amplitude RV
variations. The spectra have a S/N between 5.9 and 21.2. The resulting
radial velocities have good phase coverage and an
RMS scatter of 2.3\,\kms, comparable to the median per-point
uncertainty of 2.9\,\kms. The resulting upper limit on the mass of
the transiting companion is $\mpl < 31$\,\mjup\ at $3\sigma$ confidence.

A total of eight PFS observations were obtained for \hatcur{} between
2014 December 31 and 2017 January 13. These include seven observations
through an I$_{2}$ absorption cell, and one observation without the
cell used to construct a template spectrum for use in the RV
measurements. The observations were reduced to high-precision relative
RV measurements following \citet{butler:1996}, while spectral line
bisector spans (BSs) and their uncertainties were measured as described
by \citet{jordan:2014:hats4} and \citet{brahm:2017:ceres}. To avoid
excessive cosmic ray contamination and smearing due to changes in time
in the barycentric velocity correction, each observation was composed
of two to four exposures which were independently reduced and then
co-added. The high-precision RV and BS measurements are given in
\reftabl{rvs}, and are shown phase-folded, together with the best-fit
model, in \reffigl{rvbis}. Due to the faintness of the source, the RVs
have a median per-point uncertainty of 17\,\ms, which may be
underestimated. The residuals from the best-fit model have an RMS of
89\,\ms\ (the observations themselves have an RMS of 106\,\ms). The BS
measurements have an even larger scatter of 1.6\,\kms, limiting their
use in excluding blended eclipsing binary scenarios (such scenarios are
considered and rejected in Section~\ref{sec:blend}).

We checked the PFS observations for H$\alpha$ emission, indicative of
chromospheric activity, and found no evidence for this. If anything,
H$\alpha$ is seen in absorption in these spectra.

The surface temperature of \hatcur{} is too low to apply ZASPE
\citep{brahm:2017:zaspe}, a synthetic-template-cross-correlation-based
method to determine precise stellar atmospheric parameters, which we
have used in analyzing most of the other planetary hosts discovered by
HATSouth. For this reason we obtained a near-infrared spectrum of
\hatcur{} using the ``Astronomy Research using the Cornell Infra Red
Imaging Spectrograph'' (ARCoIRIS) instrument on the Blanco~4\,m at CTIO
\citep{arcoiris:2016}. This spectrum was used to determine \teffstar\
and \feh.

ARCoIRIS is a cross-dispersed, single-object, long-slit, near-infrared
spectrograph covering most of the wavelength range from 0.8 to 2.47
\ensuremath{\mu {\rm m} }, at a resolution of roughly 3500. ARCoIRIS
spectra can only be taken in a single setup with a fixed slit assembly
of 1\farcs1 $\times$ 28\arcsec. We observed \hatcur{} using a pair of
ABBA patterns (eight 100\,s exposures in total) interleaved with hallow
cathode lamp spectra, and using HD~1860 as a telluric standard. The
observations were carried out on UT 2016 July 15, and were reduced to
wavelength- and telluric-corrected spectra using the standard SPEX-tool
package \citep{cushing:2004, vacca:2004}. We note, that we did not
attempt to flux calibrate our spectrum as the observing conditions were
not photometric. The data reduction resulted in six extracted orders,
though we did not consider the sixth order in our analysis. Finally,
we cut out regions strongly affected by telluric lines, normalized the
spectra and removed a 2nd order polynomial fit.

In order to estimate $\teffstar$\ and \feh\ from our NIR spectrum, we
used the procedure described by \citet{newton:2015}.  These relations
were calibrated using IRTF/SpeX spectra with a resolution of
R$\sim$2,000, but ARCoIRIS has a resolution of R$\sim$3,500, therefore
we downgraded our ARCoIRS spectra to the IRTF/SpeX resolution.  In
these downgraded spectra we measured the equivalent width (EW) of some
selected lines and applied the relation from \citet{newton:2015}. 
Based on this we measure $\teffstar = \hatcurSMEteff$\,K, and \feh$=
\hatcurSMEzfeh$.

\ifthenelse{\boolean{emulateapj}}{
    \begin{figure} [ht]
}{
    \begin{figure}[ht]
}
\plotone{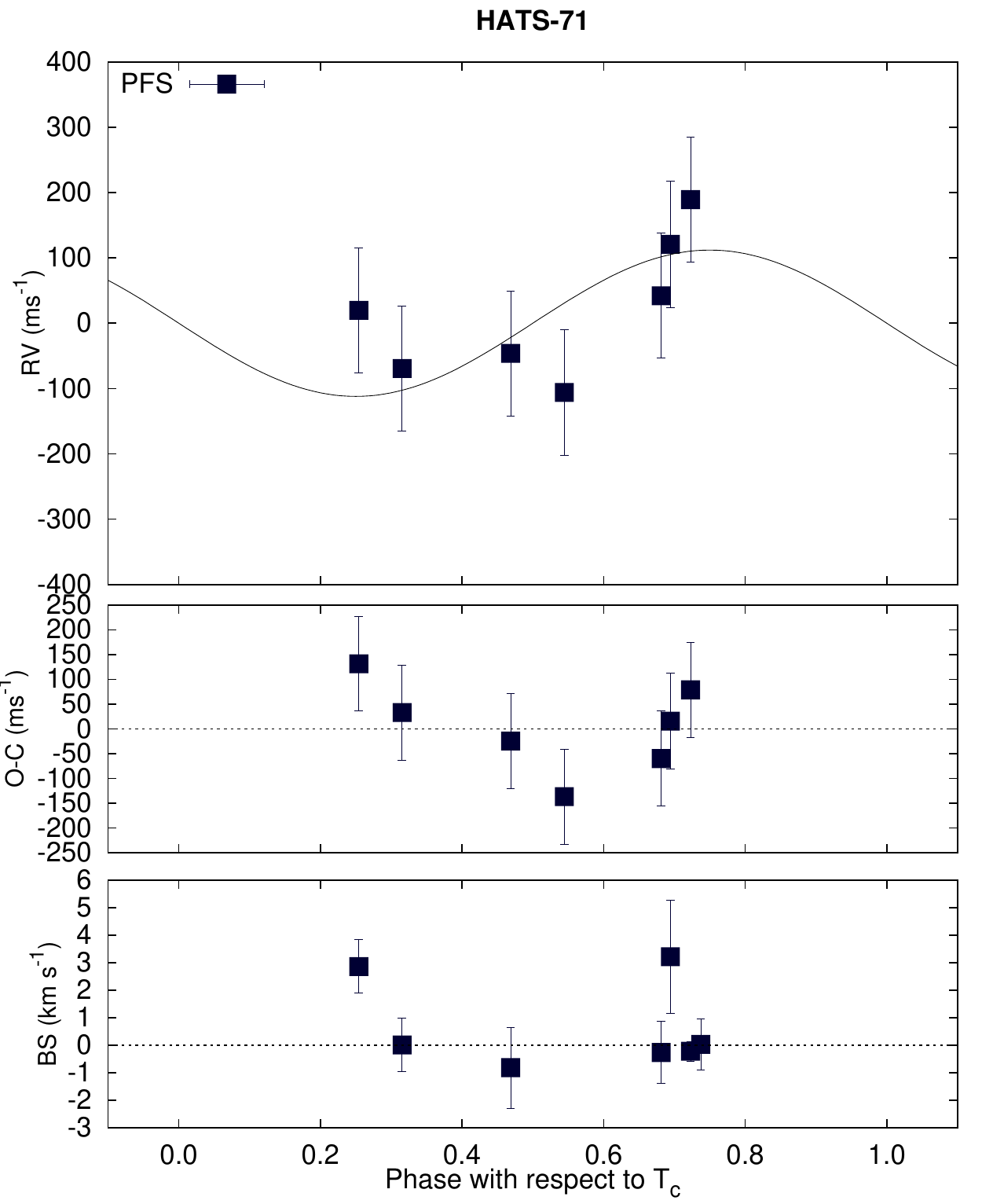}
\caption{
    Phased high-precision RV measurements from PFS for
    \hbox{\hatcur{}{}}.  {\em Top:} the phased measurements together
    with our best-fit model (see \reftabl{planetparam}).  Zero-phase
    corresponds to the time of mid-transit.  The center-of-mass
    velocity has been subtracted.  {\em Middle:} the velocity $O\!-\!C$
    residuals from the best fit.  The error bars include the jitter
    term listed in \reftabl{planetparam} added in quadrature to the
    formal errors.  {\em Bottom:} the phased bisector spans (BS).  Note
    the different vertical scales of the panels.
}
\label{fig:rvbis}
\ifthenelse{\boolean{emulateapj}}{
    \end{figure}
}{
    \end{figure}
}

\subsection{Ground-Based photometric follow-up observations}
\label{sec:phot}

\begin{figure*}[!ht]
\plotone{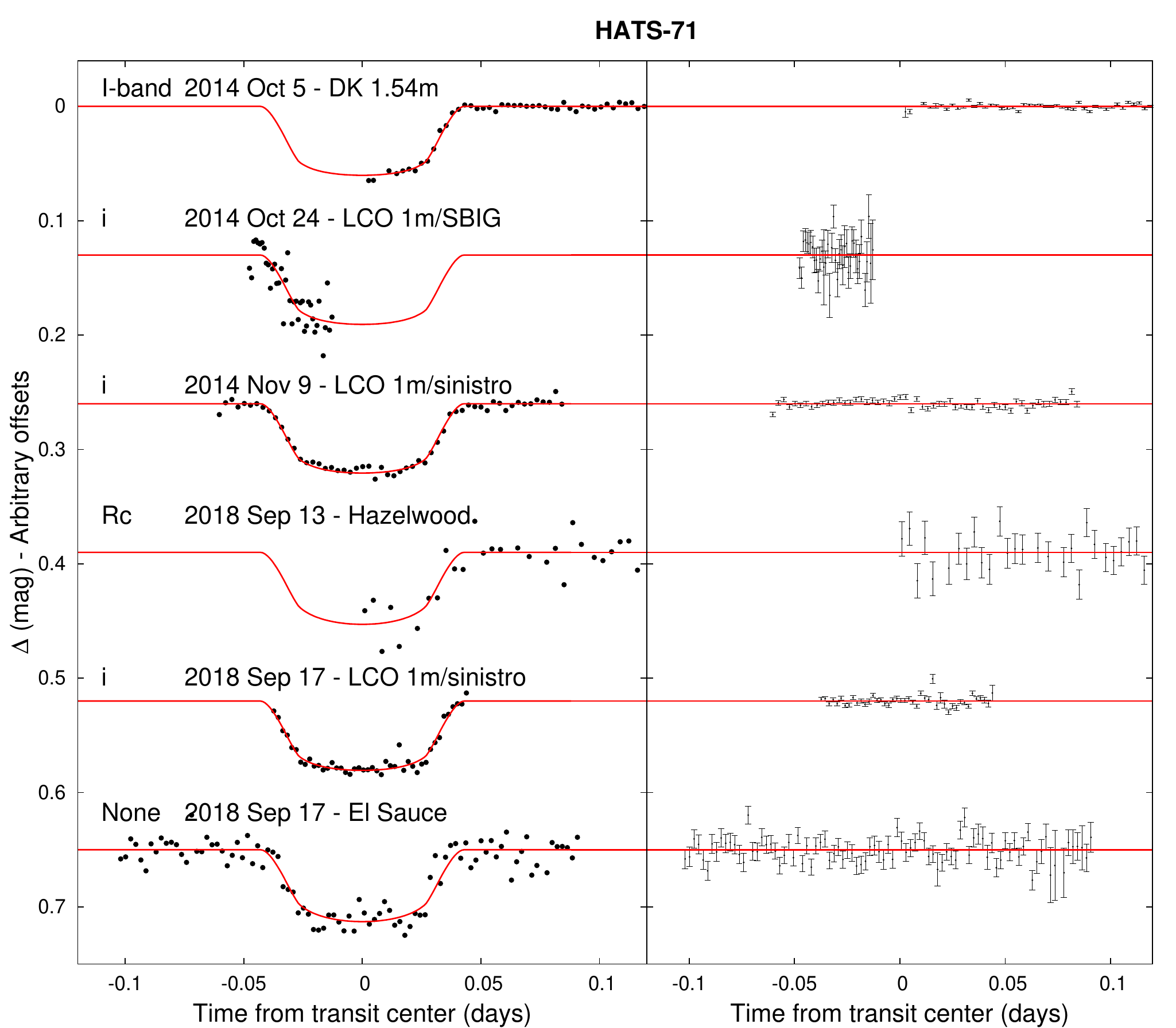}
\caption{
    Unbinned, de-trended, ground-based, follow-up transit \lcs{} for
    \hatcur{}.  The dates of the events, filters and instruments used
    are indicated.  Light curves following the first are displaced
    vertically for clarity.  Our best fit from the global modeling
    described in \refsecl{globmod} is shown by the solid lines.  The
    residuals from the best-fit model are shown on the right-hand-side
    in the same order as the original light curves.  The error bars
    represent the photon and background shot noise, plus the readout
    noise.
}
\label{fig:lc}
\end{figure*}

Follow-up higher-precision ground-based photometric transit
observations were obtained for \hatcur{} using the Danish 1.54\,m
telescope at La Silla Observatory in Chile \citep{andersen:1995}, 1\,m
telescopes from the Las Cumbres Observatory (LCOGT) network
\citep{brown:2013:lcogt}, a 0.32\,m telescope at Hazelwood Observatory
in Victoria, Australia, and a 0.36\,m telescope at El Sauce Observatory
in Chile. Three of the light curves were obtained through the {\em
TESS} Follow-up Program (TFOP) following the independent detection of
\hatcur{} as a candidate transiting planet system by the {\em TESS}
team (see Section~\ref{sec:spacephot}). All of the ground-based
follow-up light curves are shown in \reffigl{lc}, while the data are
available in \reftabl{phfu}.

An egress event was observed with the DFOSC camera on the DK~1.54\,m
telescope on the night of UT 2014 Oct 5. A total of 51 images were
collected at a median cadence of 225\,s. The observations were carried
out and reduced to a relative light curve following
\citet{rabus:2016:hats11hats12}. The residuals from the best-fit
transit model have a point-to-point RMS of 2.4\,mmag.

An ingress event was observed with the SBIG camera on one of the
LCOGT~1\,m telescopes at the South African Astronomical Observatory
(SAAO) on UT 2014 Oct 24. A total of 39 images were collected at a
median cadence of 76\,s. We also observed a full transit with the
sinistro camera on one of the LCOGT~1\,m telescopes at Cerro Tololo
Inter-American Observatory (CTIO) in Chile on UT 2014 Nov 9. A total
of 56 images were collected at a median cadence of 227\,s. These
observations were reduced to relative light curves as described in
\citet{hartman:2015:hats6}. A full transit was also observed through
the TFOP program using the sinistro camera on one of the LCOGT~1\,m
telescopes at CTIO on UT 2018 Sep 17. A total of 44 images were
collected at a median cadence of 163\,s. These data were reduced to
aperture photometry using the AstroImageJ software package
\citep[AIJ][]{collins:2013,collins:2017}. The residuals from the
best-fit transit model have a point-to-point RMS of 15\,mmag,
3.4\,mmag, and 4.6\,mmag, on each of the respective nights.

An egress event was observed on UT 2018 Sep 13 at Hazelwood
Observatory, a backyard observatory operated by Chris Stockdale in
Victoria, Australia. The observations were carried out using a 0.32\,m
Planewave CDK12 telescope and an SBIG STT-3200 CCD imager. The images
had a pixel scale of $1\farcs1$, while the average estimated PSF FWHM
on the night of the observations was $9\arcsec$. We include in the
analysis the photometry measured from 28 images collected at a median
cadence of 314\,s. Aperture photometry was performed using AIJ\@. The
residuals from the best-fit transit model have a point-to-point RMS of
15\,mmag.

A full transit was observed on UT 2018 Sep 17 at El Sauce Observatory
in Chile by Phil Evans using a 0.36\,m Planewave CDK14 telescope and a
SBIG STT1603-3 CCD imager. These images had a pixel scale of
$1\farcs47$, while the average estimate PSF FWHM on the night of the
observations was $8.2\arcsec$. A total of 90 images are included in
the analysis. The median cadence was 185\,s. Aperture photometry was
performed using AIJ\@. The residuals from the best-fit transit model
have a point-to-point RMS of 11\,mmag.

\subsection{Space-Based photometric follow-up observations}
\label{sec:spacephot}

Photometric time-series observations of \hatcur{} were carried out by
the NASA {\em TESS} mission between 2018 July 25 and 2018 August 22
(Sector 1 of the mission).  The target (\hatcurTICID{}) was selected
for observations at 2-minute cadence through the {\em TESS} Guest
Observer program\footnote{Program G011214, PI Bakos, "TESS
  Observations Of Transiting Planet Candidates From HAT"}. The data
were processed, and the source was identified as a candidate
transiting planet system (denoted~\hatcurTOIID{}) by the {\em TESS}
team following the methods described by \citet{huang:2018}. We note
that the identification of this object as a candidate by the {\em
  TESS} team was made independently of the observations described in
the previous sections.  Here we make use of the preliminary de-trended
light curve for \hatcur{} produced by the {\em TESS} Science Processing
Operations Center pipeline \citep[based on][]{jenkins:2016} which was
included in the set of {\em TESS} alerts released to the public on 2018
September 5.  Note that these Presearch Data Conditioning (PDC) light
curves have not be arbitrarily detrended, but rather have had instrumental
systematic signatures identified and removed using a multi-scale, Maximum
A Posteriori (msMAP) approach \citep{stumpe:2014,smith:2012}.
A total of 8 consecutive primary transits, and 6 epochs of
secondary eclipse are included in the light curve.  The residuals from
the best-fit model have a point-to-point RMS of 16.5\,mmag.  The light
curve is shown, together with the best-fit model, in \reffigl{tess},
while the time-series data are included in \reftabl{phfu}.

We searched for additional periodic signals in the {\em TESS} light
curve in the same manner as we did for the HATSouth data
(Section~\ref{sec:detection}). No significant signals were found with
either GLS or BLS in the {\em TESS} light curve after subtracting the
best-fit transit model for \hatcurb{}. No evidence for the
\hatcurrotper\,day photometric rotation period seen with HATSouth is observed
in the {\em TESS} data, though this is hardly surprising as this
period exceeds the duration of the {\em TESS} observations, and a
long-term linear or quadratic trend could have been filtered out by
the PDC pipeline. The highest peak in the BLS spectrum of the {\em
TESS} residuals has a period of $9.06$\,days, a depth of $3.4$\,mmag
and a signal-to-pink-noise ratio of only $5.4$.

\ifthenelse{\boolean{emulateapj}}{
    \begin{figure*}[!ht]
}{
    \begin{figure}[!ht]
}
\plotone{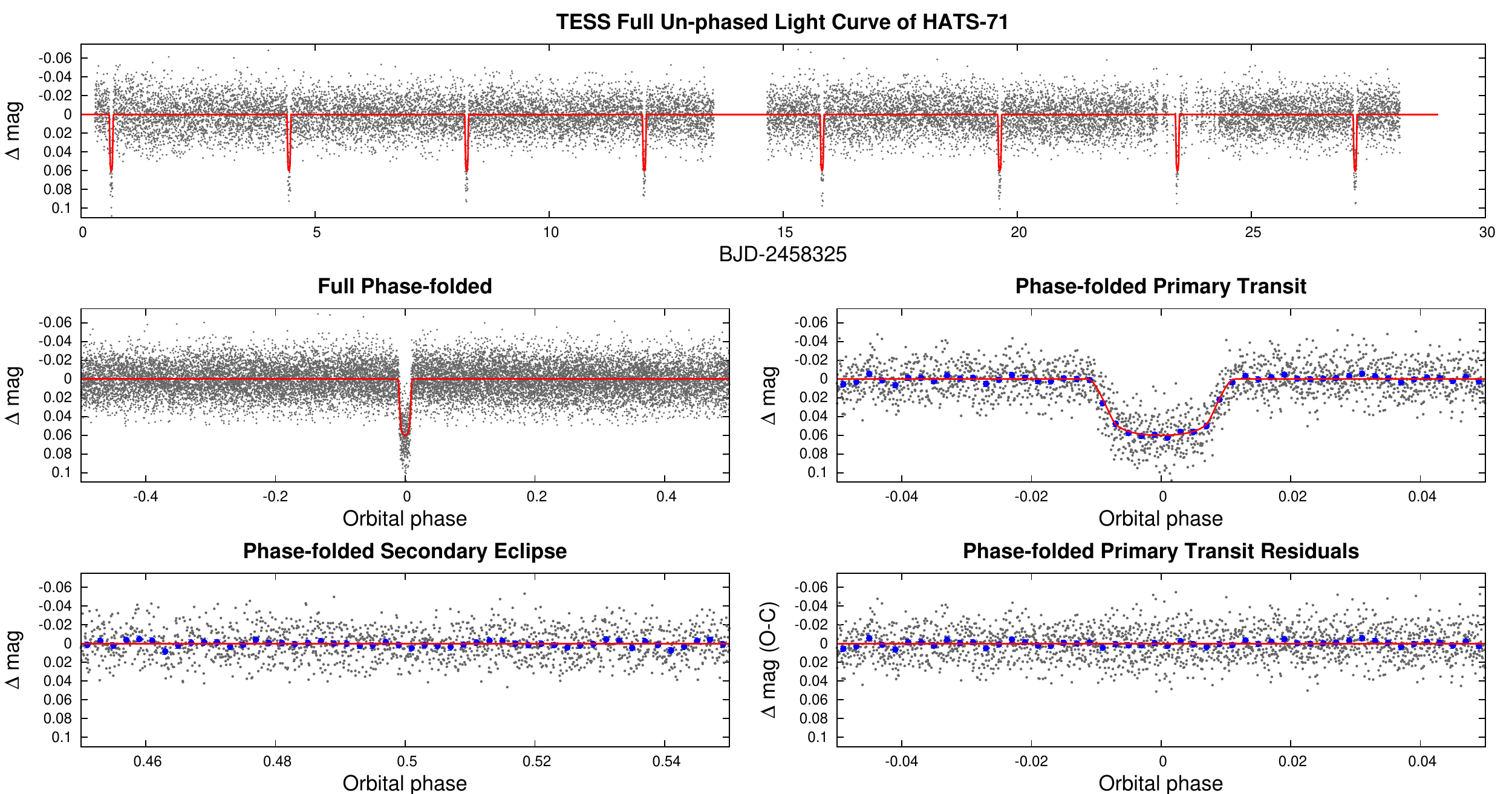}
\caption{
    {\em TESS} unbinned light curve for \hatcur{}.  We show the full
    un-phased light curve as a function of time ({\em top}), the full
    phase-folded light curve ({\em middle left}), the phase-folded
    light curve zoomed-in on the primary transit ({\em middle right}),
    the phase-folded light curve zoomed-in on the secondary eclipse
    ({\em bottom left}), and the residuals from the best-fit model,
    phase-folded and zoomed-in on the primary transit ({\em bottom
    right}).  The solid line in each panel shows the model fit to the
    light curve.  The dark filled circles show the light curve binned
    in phase with a bin size of 0.002.
\label{fig:tess}}
\ifthenelse{\boolean{emulateapj}}{
    \end{figure*}
}{
    \end{figure}
}

\ifthenelse{\boolean{emulateapj}}{
    \begin{deluxetable*}{lrrrrl}
}{
    \begin{deluxetable}{lrrrrl}
}
\tablewidth{0pc}
\tablecaption{
    Light curve data for \hatcur\label{tab:phfu}.
}
\tablehead{
    \colhead{BJD\tablenotemark{a}} & 
    \colhead{Mag\tablenotemark{b}} & 
    \colhead{\ensuremath{\sigma_{\rm Mag}}} &
    \colhead{Mag(orig)\tablenotemark{c}} & 
    \colhead{Filter} &
    \colhead{Instrument} \\
    \colhead{\hbox{~~~~(2,400,000$+$)~~~~}} & 
    \colhead{} & 
    \colhead{} &
    \colhead{} & 
    \colhead{} &
    \colhead{}
}
\startdata
\input{phfu_tab_short.tex}
\enddata
\tablenotetext{a}{
    Barycentric Julian Date computed on the TDB system with correction
	for leap seconds.
}
\tablenotetext{b}{
    The out-of-transit level has been subtracted.  For observations
    made with the HATSouth instruments (identified by ``HS'' in the
    ``Instrument'' column) these magnitudes have been corrected for
    trends using the EPD and TFA procedures applied {\em prior} to
    fitting the transit model.  This procedure may lead to an
    artificial dilution in the transit depths when used in its plain
    mode, instead of the signal reconstruction mode
    \citep{kovacs:2005:TFA}.  The blend factors for the HATSouth light
    curves are listed in Table~\ref{tab:planetparam}.  For observations
    made with follow-up instruments (anything other than ``HS'' in the
    ``Instrument'' column), the magnitudes have been corrected for a
    quadratic trend in time, and for variations correlated with up to
    three PSF shape parameters, fit simultaneously with the transit.
}
\tablenotetext{c}{
    Raw magnitude values without correction for the quadratic trend in
    time, or for trends correlated with the seeing. These are only
    reported for the follow-up observations.
}
\tablecomments{
    This table is available in a machine-readable form in the online
    journal.  A portion is shown here for guidance regarding its form
    and content.
}
\ifthenelse{\boolean{emulateapj}}{
    \end{deluxetable*}
}{
    \end{deluxetable}
}

\subsection{Search for Resolved Stellar Companions}
\label{sec:luckyimaging}

In order to detect neighboring stellar companions we obtained
$z^{\prime}$-band high-spatial-resolution lucky imaging observations
with the Astralux Sur imager \citep{hippler:2009} on the New Technology
Telescope (NTT) on the night of 2015 December 23. The observations
were reduced as in \citet{espinoza:2016:hats25hats30}, and no neighbors
were detected. The effective FWHM of the reduced image is $46.3 \pm
5.5$\,mas. Figure~\ref{fig:hatsastralux} shows the resulting $5\sigma$
contrast curve. We may exclude neighbors with $\Delta z^{\prime} <
2.5$\,mag at $0\farcs2$, and $\Delta z^{\prime} < 3.2$\,mag at
1\arcsec. We also note that there are no neighbors within 10\arcsec\
of \hatcur{} in the Gaia~DR2 catalog, based on which we rule out
neighbors with $G \la 20$\,mag down to a limiting resolution of $\sim
1\arcsec$ \citep[e.g.,][]{ziegler:2018}.

\ifthenelse{\boolean{emulateapj}}{
    \begin{figure*}[!ht]
}{
    \begin{figure}[!ht]
}
\plottwo{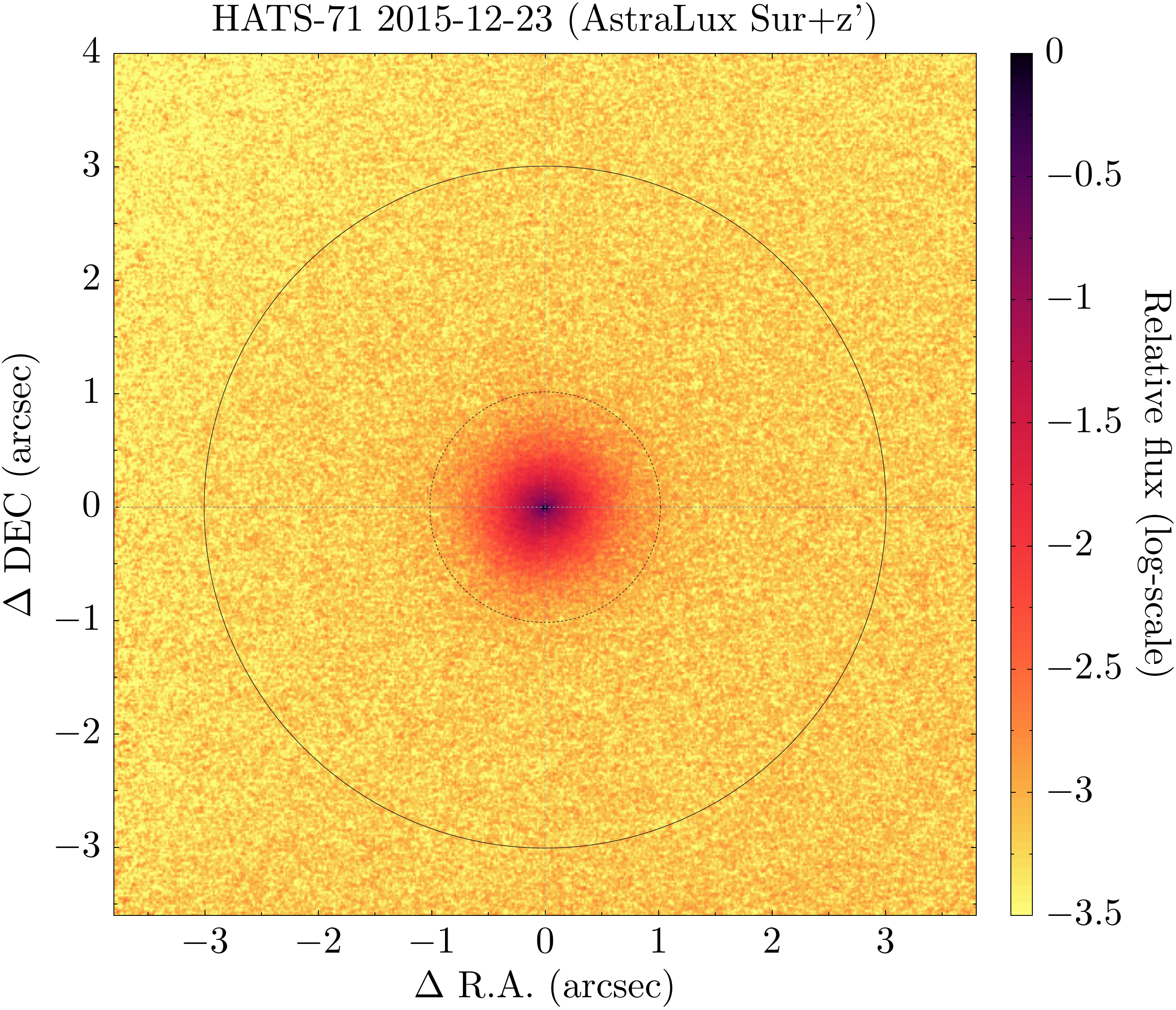}{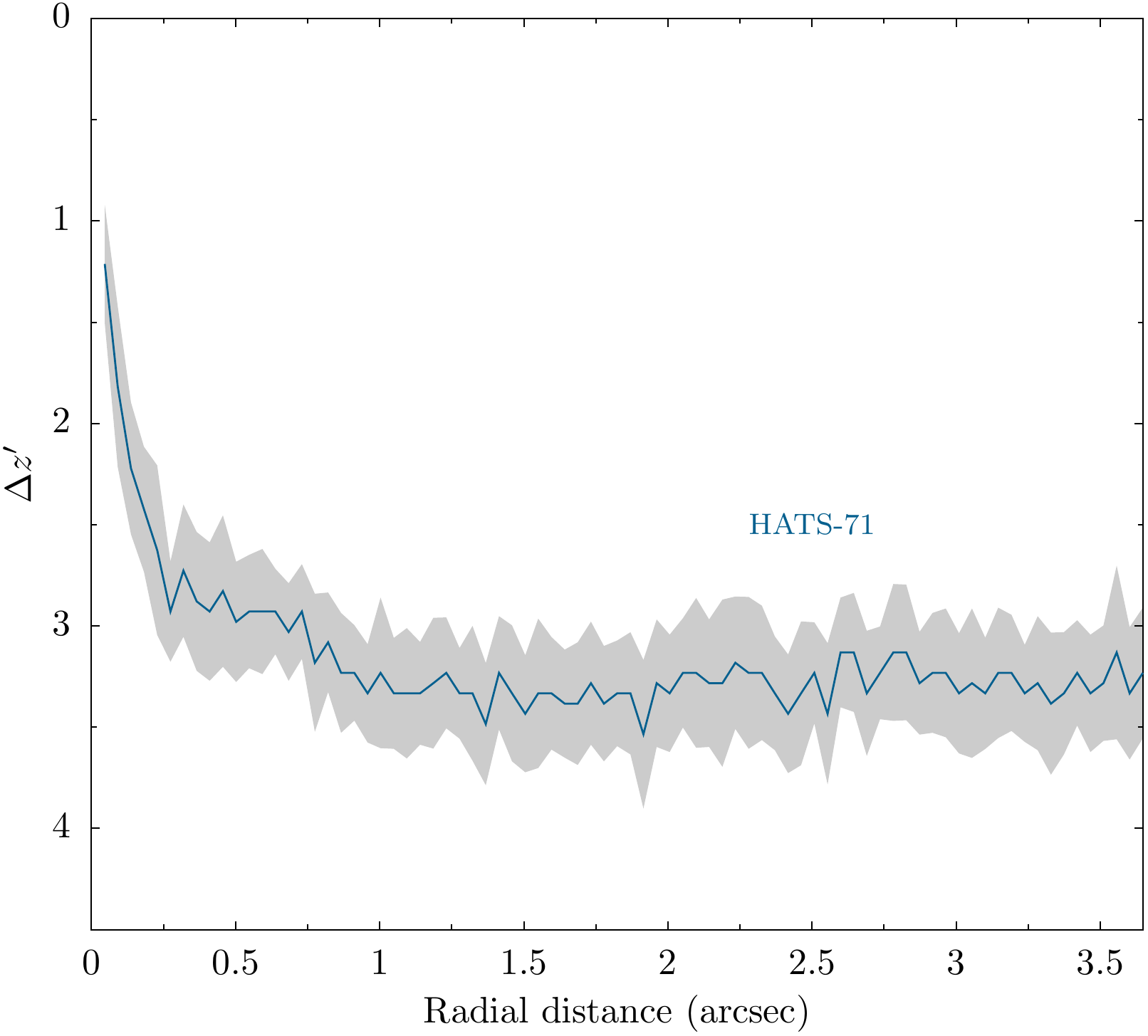}
\caption{
	{\em Left:} Astralux Sur $z^{\prime}$ image of \hatcur{} showing no
	apparent neighbors.  {\em Right:} $5\sigma$ contrast curve for
	\hatcur{} based on our Astralux Sur $z^{\prime}$ observation.  The
	gray band shows the variation in the limit in azimuth at a given
	radius.
\label{fig:hatsastralux}}
\ifthenelse{\boolean{emulateapj}}{
    \end{figure*}
}{
    \end{figure}
}

\section{Analysis}
\label{sec:analysis}

\subsection{Joint Modeling of Observations}
\label{sec:globmod}

\ifthenelse{\boolean{emulateapj}}{
    \begin{figure}[!ht]
}{
    \begin{figure}[!ht]
}
\plotone{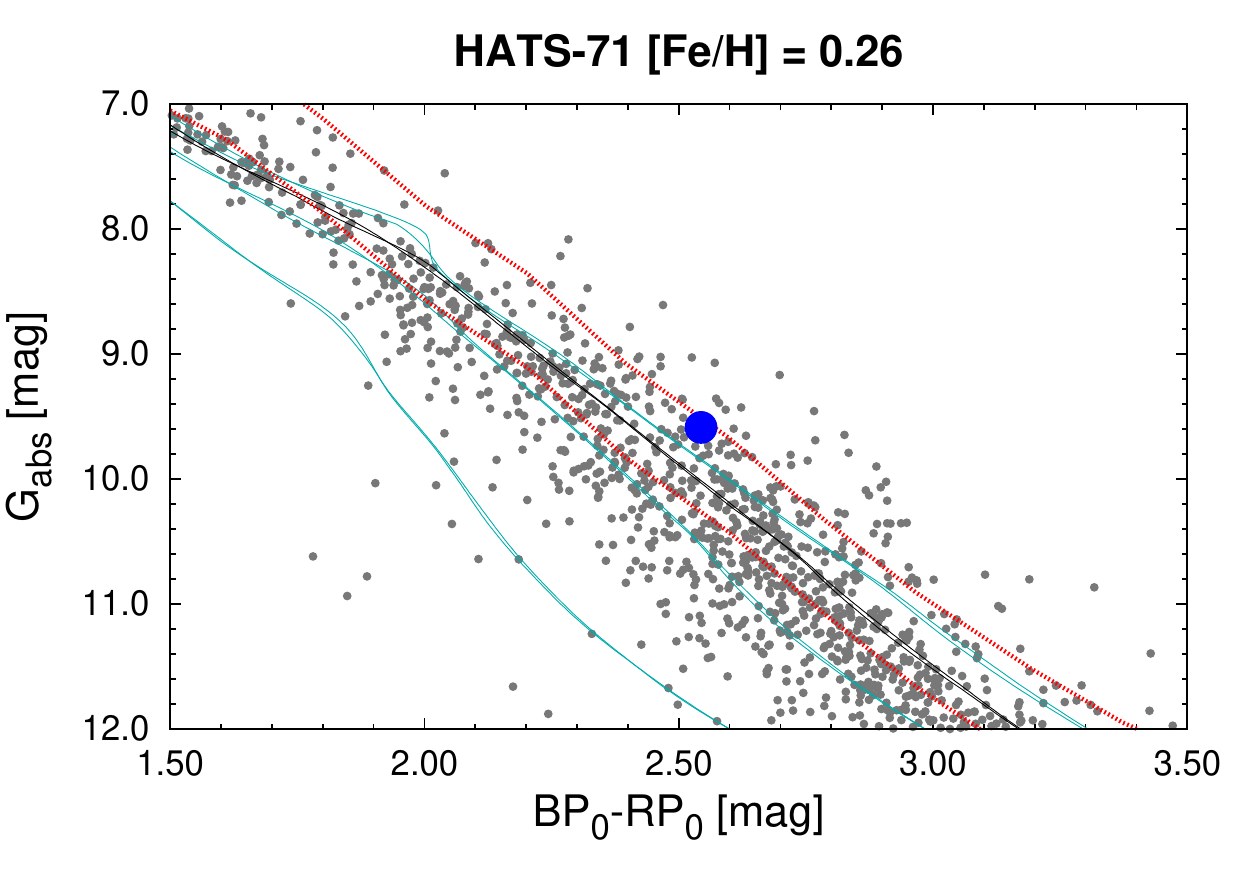}
\caption{
 Hertzsprung-Russell diagram constructed from the Gaia DR2 photometry
 corrected for distance and extinction. The blue-filled circle shows
 \hatcur{} (the uncertainties are smaller than the size of the
 circle), while the gray-filled circles show other stars in Gaia~DR2
 with $\varpi < 7$\,mas and within a $10^{\circ}\times10^{\circ}$ box
 centered on \hatcur{}. Overplotted are PARSEC model isochrones for
 metallicities of -0.5 (left set of cyan lines), 0 (middle set of cyan
 lines), +0.4414 (right set of cyan lines), and the spectroscopically
 estimated metallicity of 0.26\,dex (black lines). At each
 metallicity we show models for ages 1.0 and 5.0 and 12.0\,Gyr, though
 the difference with age at fixed metallicity is negligible at the
 scale shown here. We also show the median main sequence relation
 based on the Gaia~DR2 stars included in the plot (left red line) and
 the sequence shifted upward in magnitude by $0.753$\,mag (right red
 line; this corresponds to equal-mass binary stars with both
 components falling on the median main sequence). \hatcur{} lies near
 the upper red line, and above the +0.4414\,dex isochrones, hinting
 that it may be a unresolved binary system.
}
\label{fig:iso}
\ifthenelse{\boolean{emulateapj}}{
    \end{figure}
}{
    \end{figure}
}

We analyzed the photometric and spectroscopic observations of \hatcur{}
following \citet{hartman:2018:hats6069}. In this case we make use of
the empirical method for determining the masses and radii of the host
stars described in that paper, which is similar to the method proposed
by \citet{stassun:2018}. The method jointly fits all of the light
curves, the RV observations, the Gaia DR2 parallax, the Gaia DR2 and
2MASS broad-band photometry, and the spectroscopically determined
\teffstar\ and \feh\ (here we use the values determined from the
ARCoIRIS observations, Section~\ref{sec:obsspec}). We adopt a
Keplerian orbit to model the RV observations and \citet{mandel:2002}
light curve models in fitting the light curves, and assume fixed
quadratic limb darkening coefficients taken from \citet{claret:2004}
for $\teffstar = 3500$\,K and $\logg = 4.5$ (for the {\em TESS} light
curve we adopt the $I$-band coefficients). We used a Differential
Evolution Markov Chain Monte Carlo (DEMCMC) procedure to explore the
fitness landscape and to determine the posterior distribution of the
parameters.

This modeling allows us to directly determine the radius of the star
(making use of bolometric corrections determined from the PARSEC stellar
evolution models, \citealp{marigo:2017}; and using the MWDUST model of
\citealp{bovy:2016} to place a prior on the extinction). Combining
this with the density determined from the transits allows us to then
directly measure the mass of the star as well. In
\citet{hartman:2018:hats6069} we found that this empirical method, when
applied to the planetary systems HATS-60 through HATS-69, failed to
provide reasonably tight constraints on the stellar masses. In the
case of \hatcur{}, however, the observational constraints on the
stellar density are more stringent, allowing a significantly tighter
constraint on the stellar mass.

In carrying out the analysis we assumed a circular orbit. Note that
if the orbit is eccentric, the stellar density inferred from the light
curve would be systematically different from what we measured here,
which would in turn affect the stellar mass measurement and the
inferred planetary mass limits. A solution can be found, for example,
with $e = 0.413$ which passes nicely through the RV observations and
is consistent with the host star having a mass and radius of
$0.46$\,\msun\ and $0.45$\,\rsun, respectively, and the planet having
a mass and radius of $1.68$\,\mjup\ and $0.94$\,\rjup,
respectively. The limited number of RV observations gathered, however, prevents
us from putting a believable constraint on the eccentricity from the
data. Additional RV measurements are required, but are
expensive due to the faintness of the host star.

In fitting the DK~1.54\,m follow-up light curve we included the light
curves for 10 neighboring stars as TFA templates to account for
systematic drifts in the photometry shared by some of the comparisons
that were not well modeled by a simple function of time. For the other
ground-based follow-up light curves, where systematic variations were
less pronounced, we included only a quadratic function in time to
account for trends.

We also attempted to model the observations using the stellar
isochrone-based analysis method described by
\citet{hartman:2018:hats6069}. We found, however, that the PARSEC
theoretical model does not reproduce the high-precision measurements of
color, density and absolute magnitude that are available for \hatcur{}.

In \reffigl{iso} we show the HR diagram using the extinction- and
distance-corrected Gaia~DR2 BP$_{0}-$RP$_{0}$ and G$_{\rm abs}$
measurements. Here we show the measurements for \hatcur{} as well as
for all stars in the Gaia~DR2 catalog in a $10^{\circ}\times10^{\circ}$
box centered on \hatcur{} with parallax $\varpi > 7$\,mas,
$\sigma_{\varpi} < 0.2$\,mas, and BP, RP, and G all measured to greater
than 10$\sigma$ confidence, and with $1.5 < $BP$-$RP$_{0} < 3.5$ and
$7.0 < $G$_{\rm abs} < 12.0$. We also show theoretical PARSEC
isochrones for a range of ages and metallicities, the median main
sequence relation based on the selected stars from the Gaia~DR2 sample,
and the median main sequence shifted upward in magnitude by
$0.753$\,mag (corresponding to equal-mass binary stars with both
components falling on the median main sequence). As is apparent,
\hatcur{} falls above the highest metallicity theoretical relation
calculated, and near the equal-mass binary sequence. This provides
suggestive evidence that \hatcur{} may be an unresolved binary star
system, though we caution that there is no other spectroscopic or
imaging evidence for such a companion. We consider how the inferred
planetary and stellar parameters would change if there is an unresolved
stellar companion in \refsecl{blend}.

Previous work has shown that rapidly rotating, magnetically active M
dwarfs often have cooler surface temperatures and larger radii than
predicted by theoretical stellar evolution models (e.g., see the recent
work by \citealp{jaehnig:2018} and \citealp{somers:2017} investigating
the inflation of M dwarfs in the Hyades and Pleiades; see also
references therein for a rich literature on this topic). \hatcur{},
however, does not exhibit H$\alpha$ emission typical of magnetically
active M dwarfs, and its measured photometric rotation period of
\hatcurrotper\,days (Section~\ref{sec:detection}) is substantially
longer than the periods of M dwarf stars for which radius inflation is
typically observed ($P_{\rm rot} \la 10$\,days).

The measured astrometric, spectroscopic and photometric parameters of
\hatcur{} are collected in \reftabl{stellarobserved}. 
\reftabl{stellarderived} gives the stellar parameters that are derived
through the modelling discussed in this Section, while
\reftabl{planetparam} gives the planetary parameters derived through
this modeling. The parameters listed under the ``Single Star'' columns
in each table are those derived here under the assumption that
\hatcur{} is a single star without a stellar binary companion.

We find that, thanks to Gaia~DR2, the star \hatcur{} has a tightly
constrained radius of \hatcurISOrlong\,\rsun. This, combined with the
measured bulk stellar density (from the transits) of
\hatcurLCrho{}\,\gcmc, gives a stellar mass of \hatcurISOmlong\,\msun. 
For comparison, using the \citet{delfosse:2000} mass--M$_{K}$ relation
gives an estimated stellar mass of $0.455$\,\msun, while using the
\citet{benedict:2016} mass--M$_{K}$ relation gives an estimated stellar
mass of $0.50$\,\msun, consistent with the value coming from Gaia~DR2
and the mean density estimate. 
%

We find that the planet \hatcurb{} has a radius of
\hatcurPPrlong\,\rjup.  Due to the faintness of the source we are
unable to determine the mass of the planet with greater than $2\sigma$
confidence.  Our modeling yields a mass of \hatcurPPmlong\,\mjup, with
a 95\% confidence upper limit of $\mpl < 0.81$\,\mjup.  The planet has
an estimated equilibrium temperature (assuming full redistribution of
heat and zero albedo) of \hatcurPPteff\,K.

The 89\,\ms\ scatter in the PFS RV residuals is significantly larger
than the median per-point uncertainty of 17\,\ms. Given the limited
number of RVs obtained we cannot say whether this is due to the planet
having a significant eccentricity, stellar activity, additional planets
in the system, or our underestimating the uncertainties in these low
S/N spectra. In modeling the data we incorporated a jitter term, which
we added in quadrature to the formal uncertainties, and varied in the
fit. We find a jitter of \hatcurRVjitter\,\ms\ is needed to explain
the excess scatter. If the orbit is eccentric, the jitter could be as
low as $37$\,\ms.

\subsection{Blend Analysis}
\label{sec:blend}

In order to rule out the possibility that \hatcur{} is a blended
stellar eclipsing binary system, we carried out a blend analysis of the
photometric data following \citet{hartman:2018:hats6069}.  In this
analysis we model the photometric and spectroscopic observations of
\hatcur{} under four different scenarios: a single star with a planet
(referred to as the H-p model following the nomenclature from
\citealp{hartman:2009:hat12}), a hierarchical triple star system where
the two fainter stars form an eclipsing binary (referred to as the
H,S-s model), a blend between a bright foreground star and a fainter
background eclipsing binary star system (referred to as the H,S-s$_{\rm
BGEB}$ model), and a bright star with a transiting planet and a fainter
unresolved stellar companion (referred to as the H-p,s model).

We find that the best-fitting model is the H-p,s model which yields
$\Delta \chi^2 = -345$, $-278$ and $-657$ compared to the best-fit H-p
model, H,S-s$_{\rm BGEB}$ and H,S-s models, respectively. The H,S-s
model is strongly disfavored, however the H,S-s$_{\rm BGEB}$ provides
a better fit to the data modeled in this analysis than the H-p model.
As noted in Section~\ref{sec:globmod} the PARSEC models do not
reproduce the combined high-precision measurements of color, density
and absolute magnitude that are available for \hatcur\ assuming a
single star, so it is perhaps not surprising that the H,S-s$_{\rm
  BGEB}$ model can provide a better fit than the H-p model.  The
best-fit H,S-s$_{\rm BGEB}$ model consists of a 0.42\,\msun\ foreground
star blended with a $0.44+0.12$\,\msun\ eclipsing binary at a distance
modulus that is 0.65\,mag greater than the foreground star, and we find
that the primary star in the background binary can be at most only
1\,mag fainter in apparent brightness than the foreground star.  Based
on the Astralux Sur imaging (Section~\ref{sec:luckyimaging}) the
projected separation between the foreground star and the background
binary would have to be $\lesssim 0\farcs05$.  This H,S-s$_{\rm BGEB}$
model still fails to fit the observations to within the uncertainties,
yielding, for example, a predicted parallax of 6.93\,mas for the
foreground star which differs from the measured value of
\hatcurCCparallax\,mas by $4\sigma$.  What is more, we find that all of
the H,S-s$_{\rm BGEB}$ blend models which fit the observations as well
as or better than the H-p model (i.e., have $\Delta \chi^2 < 25$
compared to the H-p model) predict a significantly larger RV variation
measured from the composite spectrum than observed (with RMS ranging
from 660\,\ms\ to 1.2\,\kms).  Based on these factors we consider both
the H,S-s and H,S-s$_{\rm BGEB}$ models excluded, and conclude that
\hatcur{} is a confirmed transiting planet system.

Because the H-p,s model provides a significantly better fit to the data
than the H-p model, we also list in \reftabl{stellarderived} and
\reftabl{planetparam} the stellar parameters (for both the primary and
secondary stars) and the planetary parameters for the H-p,s model
derived from a DEMCMC analysis.  Based on this modeling, we find that
the planetary host star \hatcur{}A has a mass of
$\hatcurISOmlonghpsmodel{}$\,\msun, and a radius of
$\hatcurISOrlonghpsmodel{}$\,\rsun, while the unresolved binary star
\hatcur{}B has a mass of $\hatcurISOmlongBhpsmodel{}$\,\msun\ and a
radius of $\hatcurISOrlongBhpsmodel{}$\,\rsun.  The planet has a radius
of $\hatcurPPrlonghpsmodel{}$\,\rjup\ and a poorly determined mass of
$\hatcurPPmlonghpsmodel{}$\,\mjup\ (95\% confidence upper limit of
$\hatcurPPmtwosiglimhpsmodel{}$\,\mjup).  Here we do not incorporate
the RV observations directly into the modeling in this case, but
instead determine an approximate scaling factor of $1.16 \pm 0.23$,
which we apply to the value of $K$ as determined in \refsecl{globmod}
for the single star modeling to account for the effective dilution in
the measured orbital variation of the primary star due to the
non-varying spectral features contributed by the secondary star.  This
scaling factor is calculated by simulating blended spectral
cross-correlation functions in the same manner as done in ruling out
the H,S-s$_{\rm BGEB}$ model, and we conservatively assume a 20\%
uncertainty.  We then re-calculate all parameters that depend on K
after applying this scaling.

We also find that \hatcur{}B would have $\Delta G = 2.05$\,mag, $\Delta
z = 1.77$\,mag compared to \hatcur{}A.  Based on the Astralux Sur
observations (Section~\ref{sec:luckyimaging}) the two stars would have
to be separated by less than $0\farcs1$, implying a projected physical
separation of less than 14\,AU.  Note that we also checked whether
there was enough proper motion for HATS-71 to have moved between
archival images, but the proper motion was too small to reveal anything
by blinking the UK Schmidt image (1997) and the DK 1.5m telescope image
(2014).  The slight over-luminosity of \hatcur{} could also be caused
by being a very young M-dwarf.  However, a query with BANYAN Sigma
\citep{gagne:2018} yields no matches, so the star is unlikely to be the
member of a young association.


\section{Discussion}
\label{sec:discussion}

The discovery of \hatcurb{} demonstrates that, at least in some cases,
Jupiter-sized planets are able to form and migrate around stars with
masses as low as \hatcur{} ($\hatcurISOmlong{}$\,\msun). It remains to
be seen whether such planets occur with the same frequency as they do
around solar-type stars (i.e.~0.43$\pm0.05$\%: \citet{fressin:2013}),
or if giant planet formation is rarer around low mass stars as
predicted by core accretion theory
\citep[e.g.][]{laughlin:2004,liu:2016}. Figure~\ref{fig:massmass}
shows giant planet masses as a function of host star mass, for systems
with measured planetary masses. \hatcurb{} is the giant planet with
the lowest host star mass that has been discovered to-date. The
sparsity of systems with host masses $<$0.5\,\msun\ is apparent from
Figure~\ref{fig:massmass}, although this may just be an reflection of
the fact that most of the surveys contributing to the discoveries shown
did not monitor sufficient numbers of low mass stars. Over the next
two years of HATSouth and \textit{TESS} discoveries, we should gain a
better statistical understanding of these systems.

The deep transits that these systems present makes photometric
detection relatively robust in both the HATSouth and \textit{TESS}
survey data. Indeed, the 4.7\% transit for \hatcurb\ makes
this the deepest transit observed by a hot Jupiter (as defined in the
Introduction). In \reffig{perdepth} we show the transit depths of these planets
as a function of period, where the depths were calculated from the
$\rpl$ planetary radius, $\rstar$ stellar radius, $b$ impact parameter,
$e$ eccentricity and $\omega$ argument of periastron of the orbit
(whenever available), also taking into account the grazing nature of
some orbits. The second and third deepest transits are 
Qatar-4b \citep[3.4\%; ][]{alsubai:2017:qatar4b}, and 
HATS-6b, \citep[3.3\%; ][]{hartman:2015:hats6}. 

\ifthenelse{\boolean{emulateapj}}{
    \begin{figure}[!ht]
}{
    \begin{figure}[!ht]
}
\plotone{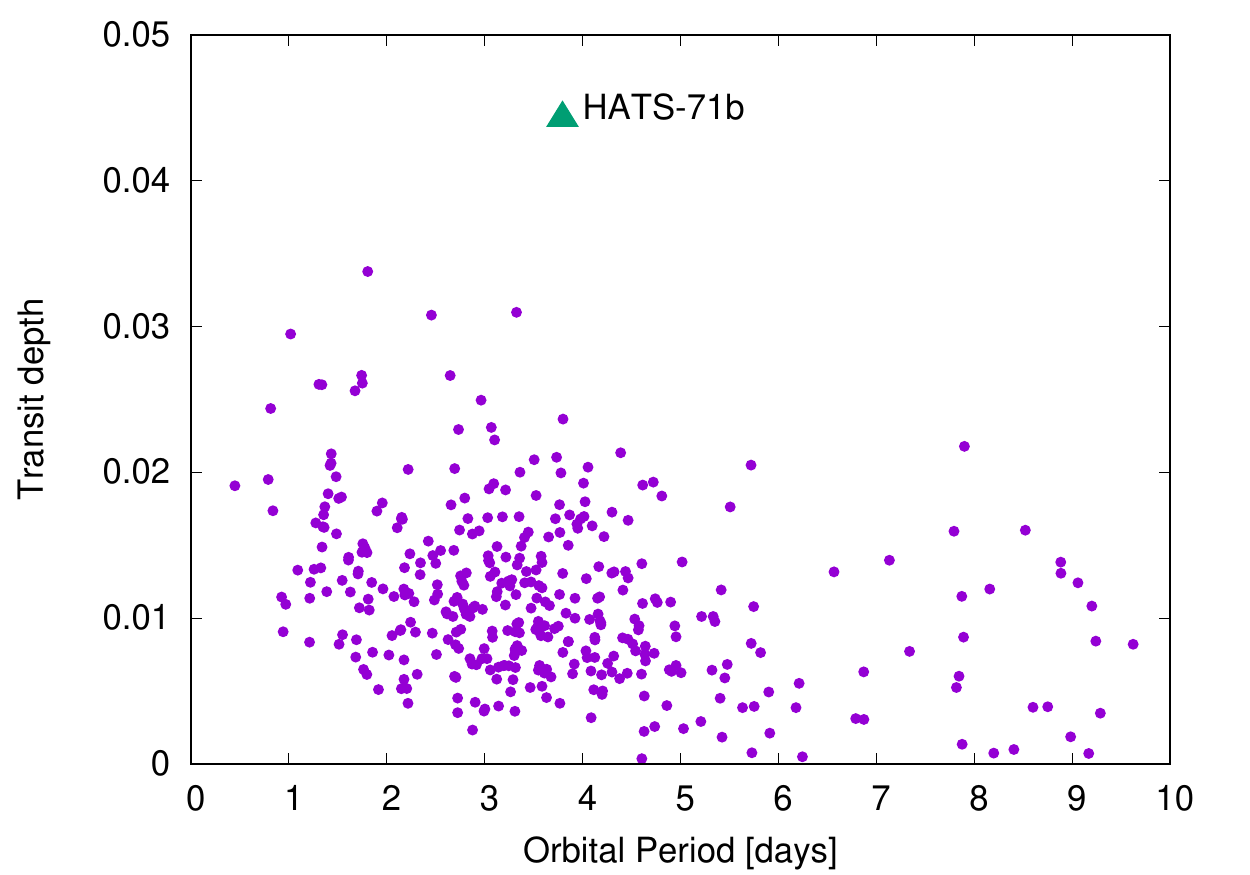}
\caption{
	Transit depth as a function of orbital period for hot Jupiters.
	The depth was calculated from the
	$\rpl$ planetary radius, $\rstar$ stellar radius, $b$ impact parameter,
	$e$ eccentricity and $\omega$ argument of periastron of the orbit
	(whenever available), also taking into account the grazing nature of
	some orbits. Data was taken from \url{exoplanet.eu}. 
}
\label{fig:perdepth}
\ifthenelse{\boolean{emulateapj}}{
    \end{figure}
}{
    \end{figure}
}

However, radial
velocity follow-up is extremely challenging since such stars are
generally faint at visible wavelengths where most high precision
spectrographs operate. The spectrum of these stars may also be less
amendable to measuring precise radial velocity variations, as they are
dominated by broader molecular absorption features rather than the
narrow metal lines in solar-type stars (see Figure~\ref{fig:wifes}). A
new generation of stable IR spectrographs will measure precise radial
velocities in order to search for planets orbiting M-dwarfs, and these
include CARMENES \citep{carmenes:2014}, SPIROU \citep{spirou:2014}, IRD
\citep{ird:2014}, HPF \citep{hpf:2018}, NIRPS \citep{nirps:2017} and
GIARPS \citep{claudi:2018}.

This may provide another avenue for radial velocity follow-up of
transiting giant planets orbiting M-dwarfs. However we note that for
mid M-dwarfs such as \hatcur{}, optical spectroscopy will probably
remain the best source of high precision radial velocities. For the
CARMENES spectrograph, which hosts both an optical and IR arm, it
appears that the radial velocity precision is still higher in the
optical wavelengths until spectral types of M8 or later
\citep{reiners:2018}.

The deep transits will facilitate atmospheric characterization of the
planet using transmission spectroscopy. We estimate that the
transmission signature could be anywhere from 300\,ppm to 700\,ppm,
assuming the cloud properties of hot Jupiters around M stars are
similar to those around F, G and K stars. Atmospheric characterization
might be used instead of radial velocities to get the mass of the
planet via MassSpec \citep{wit:2013}, although note the ambiguities
detailed in \citet{batalha:2017}.

\hatcur{} was observed by the \textit{TESS} spacecraft with 2 minute
cadence as a candidate from the HATSouth Guest Observer Program
(GO11214; PI Bakos). Due to the high precision ground-based light
curves that had already been obtained in 2014 using 1\,m-class
telescopes (see Section~\ref{sec:phot}), the addition of the
\textit{TESS} light curve did not have a significant impact on
parameters such as the planetary radius or the orbital ephemerides. 
However the \textit{TESS} light curve did contain the best photometry
available at phase 0.5, which allowed us to rule out a secondary
eclipse with much higher confidence. With many hundreds of transiting
planet candidates, follow-up photometry that covers both the primary
transit and any possible secondary eclipse is a time consuming and
resource intensive task. The use of \textit{TESS} light curves to help
confirm existing candidates is therefore an obvious synergy between
HATSouth and \textit{TESS}, and this method will continue to be adopted
for future \textit{TESS} sectors.

\ifthenelse{\boolean{emulateapj}}{
    \begin{figure}[!ht]
}{
    \begin{figure}[!ht]
}
\plotone{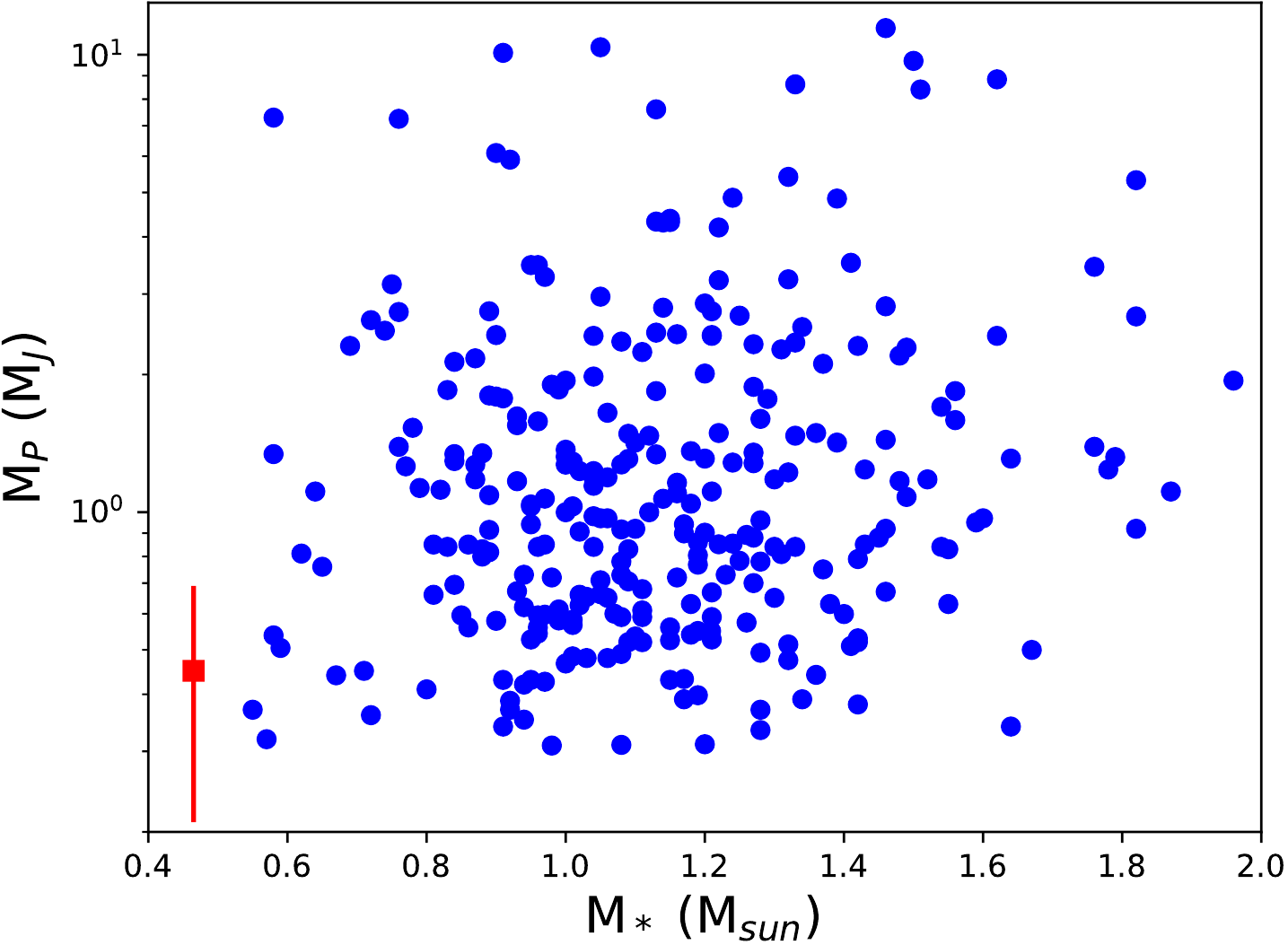}
\caption{
	Planet mass as a function of host star mass for all known giant
	(\mpl$>$0.3\,\mjup) planets with measured masses and radii (blue
	circles) and for \hatcurb{} (red square with errorbars).  Data from
	NASA Exoplanet Archive as of 2018 October 4.
}
\label{fig:massmass}
\ifthenelse{\boolean{emulateapj}}{
    \end{figure}
}{
    \end{figure}
}


\acknowledgements 

Development of the HATSouth project was funded by NSF MRI grant
NSF/AST-0723074, operations have been supported by NASA grants
NNX09AB29G, NNX12AH91H, and NNX17AB61G, and follow-up observations have
received partial support from grant NSF/AST-1108686. G.\'A.B wishes to
thank Konkoly Observatory of the Hungarian Academy of Sciences for
their warm hospitality during numerous visits during the past years, in
particular the Distinguished Guest Fellow program. A.J.\
acknowledges support from FONDECYT project 1171208, BASAL CATA
AFB-170002, and project IC120009 ``Millennium Institute of Astrophysics
(MAS)'' of the Millenium Science Initiative, Chilean Ministry of
Economy. N.E.\ is supported by CONICYT-PCHA/Doctorado Nacional. R.B.\
acknowledges support from FONDECYT Post-doctoral Fellowship Project No. 
3180246. N.E.\ acknowledges support from project IC120009 ``Millenium
Institute of Astrophysics (MAS)'' of the Millennium Science Initiative,
Chilean Ministry of Economy. L.M.\ acknowledges support from the
Italian Minister of Instruction, University and Research (MIUR) through
FFABR 2017 fund. L.M.\ acknowledges support from the University of
Rome Tor Vergata through ``Mission: Sustainability 2016'' fund. V.S.\
acknowledges support form BASAL CATA PFB-06. A.V.~is supported by the
NSF Graduate Research Fellowship, Grant No. DGE 1144152.

Based on observations at Cerro Tololo Inter-American Observatory,
National Optical Astronomy Observatory (NOAO Prop.~ID 2016A/CN-615,
2016B-CN0908, 2017A-C79, 2017B-0909, 2018A-CN46/908; PI: Rabus), which
is operated by the Association of Universities for Research in
Astronomy (AURA) under a cooperative agreement with the National
Science Foundation. M.R.~acknowledges support from CONICYT project
Basal AFB-170002.

This paper also makes use of observations from the LCOGT network. Some
of this time was awarded by NOAO. We acknowledge the use of the AAVSO
Photometric All-Sky Survey (APASS), funded by the Robert Martin Ayers
Sciences Fund, and the SIMBAD database, operated at CDS, Strasbourg,
France.
This work has made use of data from the European Space Agency (ESA)
mission {\it Gaia} (\url{https://www.cosmos.esa.int/gaia}), processed
by the {\it Gaia} Data Processing and Analysis Consortium (DPAC,
\url{https://www.cosmos.esa.int/web/gaia/dpac/consortium}). Funding
for the DPAC has been provided by national institutions, in particular
the institutions participating in the {\it Gaia} Multilateral
Agreement. 
This research has made use of the NASA Exoplanet Archive,
which is
operated by the California Institute of Technology, under contract with
the National Aeronautics and Space Administration under the Exoplanet
Exploration Program. We acknowledge the use of TESS Alert data, which
is currently in a beta test phase, from pipelines at the TESS Science
Office and at the TESS Science Processing Operations Center. This
paper includes data collected by the TESS mission, which are publicly
available from the Mikulski Archive for Space Telescopes (MAST).
Finally, G.\'A.B.~wishes to thank Princeton's AST205 class for all the
inspiration they gave during the fall semester of 2018. 


\facilities{HATSouth, ATT (WiFeS), Magellan:Clay (PFS), Blanco
(ARCoIRIS), Danish 1.54m Telescope (DFOSC), LCOGT, NTT (Astralux Sur),
TESS, Gaia, Exoplanet Archive}

\software{FITSH \citep{pal:2012}, BLS \citep{kovacs:2002:BLS}, VARTOOLS
\citep{hartman:2016:vartools}, CERES \citep{brahm:2017:ceres},
AstroImageJ \citep{collins:2013,collins:2017}, SPEX-tool
\citep{cushing:2004,vacca:2004}, SExtractor \citep{bertin:1996},
Astrometry.net \citep{lang:2010}, MWDUST \citep{bovy:2016}}


\bibliographystyle{aasjournal}
\bibliography{hatsbib}

\ifthenelse{\boolean{emulateapj}}{
    \begin{deluxetable*}{lcl}
}{
    \begin{deluxetable}{lcl}
}
\tablewidth{0pc}
\tabletypesize{\tiny}
\tablecaption{
    Astrometric, Spectroscopic and Photometric parameters for \hatcur{}
    \label{tab:stellarobserved}
}
\tablehead{
    \multicolumn{1}{c}{~~~~~~~~Parameter~~~~~~~~} &
    \multicolumn{1}{c}{Value}                     &
    \multicolumn{1}{c}{Source}
}
\startdata
\noalign{\vspace{-3pt}}
\sidehead{Astrometric properties and cross-identifications}
~~~~2MASS-ID\dotfill                     & \hatcurCCtwomassshort{}        & \\                            
~~~~TIC-ID\dotfill                       & \hatcurTICID                   & \\                            
~~~~Gaia~DR2-ID\dotfill                  & \hatcurCCgaiadrtwoshort{}      & \\                            
~~~~R.A. (J2000)\dotfill                 & \hatcurCCra{}                  & Gaia DR2\\                    
~~~~Dec. (J2000)\dotfill                 & \hatcurCCdec{}                 & Gaia DR2\\                    
~~~~$\mu_{\rm R.A.}$ (\masy)             & \hatcurCCpmra{}                & Gaia DR2\\                    
~~~~$\mu_{\rm Dec.}$ (\masy)             & \hatcurCCpmdec{}               & Gaia DR2\\                    
~~~~parallax (mas)                       & \hatcurCCparallax{}            & Gaia DR2\\                    
\sidehead{Spectroscopic properties}
~~~~$\teffstar$ (K)\dotfill              & \hatcurSMEteff{}               & ARCoIRIS\tablenotemark{a}\\   
~~~~$\feh$\dotfill                       & \hatcurSMEzfeh{}               & ARCoIRIS               \\     
~~~~$\gamma_{\rm RV}$ (\ms)\dotfill      & \hatcurRVgammaabs{}            & WiFeS\tablenotemark{b}  \\    
\sidehead{Photometric properties}
~~~~$P_{\rm rot}$ (days)\dotfill                  & \hatcurrotper{}         & HATSouth \\
~~~~$G$ (mag)\tablenotemark{c}\dotfill   & \hatcurCCgaiamG{}              & Gaia DR2 \\                   
~~~~$BP$ (mag)\tablenotemark{c}\dotfill  & \hatcurCCgaiamBP{}             & Gaia DR2 \\                   
~~~~$RP$ (mag)\tablenotemark{c}\dotfill  & \hatcurCCgaiamRP{}             & Gaia DR2 \\                   
~~~~$g$ (mag)\dotfill                    & \hatcurCCtassmg{}              & APASS\tablenotemark{d} \\     
~~~~$r$ (mag)\dotfill                    & \hatcurCCtassmr{}              & APASS\tablenotemark{d} \\     
~~~~$i$ (mag)\dotfill                    & \hatcurCCtassmi{}              & APASS\tablenotemark{d} \\     
~~~~$J$ (mag)\dotfill                    & \hatcurCCtwomassJmag{}         & 2MASS           \\            
~~~~$H$ (mag)\dotfill                    & \hatcurCCtwomassHmag{}         & 2MASS           \\            
~~~~$K_s$ (mag)\dotfill                  & \hatcurCCtwomassKmag{}         & 2MASS           \\            
\enddata
\tablenotetext{a}{
	``Astronomy Research using the Cornell Infra Red Imaging
	Spectrograph'' (ARCoIRIS) instrument on the Blanco~4\,m at CTIO
	\citep{arcoiris:2016}. 
}
\tablenotetext{b}{
    The error on $\gamma_{\rm RV}$ is determined from the
    orbital fit to the RV measurements, and does not include the
    systematic uncertainty in transforming the velocities to the IAU
    standard system. The velocities have not been corrected for
    gravitational redshifts.
} 
\tablenotetext{c}{
    The listed uncertainties for the Gaia DR2 photometry are taken from
	the catalog.  For the analysis we assume additional systematic
	uncertainties of 0.002\,mag, 0.005\,mag and 0.003\,mag for the G,
	BP and RP bands, respectively.
}
\tablenotetext{d}{
    From APASS DR6 for as
    listed in the UCAC 4 catalog \citep{zacharias:2013:ucac4}.  
}
\ifthenelse{\boolean{emulateapj}}{
    \end{deluxetable*}
}{
    \end{deluxetable}
}

\ifthenelse{\boolean{emulateapj}}{
    \begin{deluxetable*}{lcc}
}{
    \begin{deluxetable}{lcc}
}
\tablewidth{0pc}
\tabletypesize{\footnotesize}
\tablecaption{
    Derived stellar parameters for \hatcur{} system
    \label{tab:stellarderived}
}
\tablehead{
    \multicolumn{1}{c}{~~~~~~~~Parameter~~~~~~~~} &
    \multicolumn{1}{c}{Value} &
    \multicolumn{1}{c}{Value} \\
    \multicolumn{1}{c}{} &
    \multicolumn{1}{c}{Single Star} &
    \multicolumn{1}{c}{Binary Star}
}
\startdata
\sidehead{Planet Hosting Star \hatcur{}A}
~~~~$\mstar$ ($\msun$)\dotfill           & \hatcurISOmlong{}              & \hatcurISOmlonghpsmodel{}   \\
~~~~$\rstar$ ($\rsun$)\dotfill           & \hatcurISOrlong{}              & \hatcurISOrlonghpsmodel{}   \\
~~~~$\loggstar$ (cgs)\dotfill            & \hatcurISOlogg{}               & \hatcurISOlogghpsmodel{}    \\ 
~~~~$\rhostar$ (\gcmc)\dotfill           & \hatcurLCrho{}                 & \hatcurLCrhohpsmodel{}      \\   
~~~~$\lstar$ ($\lsun$)\dotfill           & \hatcurISOlum{}                & \hatcurISOlumhpsmodel{}     \\ 
~~~~$\teffstar$ (K)\dotfill              & \hatcurISOteff{}               & \hatcurISOteffhpsmodel{}    \\   
~~~~\feh\ (dex)\dotfill                  & \hatcurISOfeh{}                & \hatcurISOfehhpsmodel{}     \\    
~~~~Age (Gyr)\dotfill                    & $\cdots$                       & \hatcurISOagehpsmodel{}     \\    
~~~~$A_{V}$ (mag)\dotfill                & \hatcurXAv{}                   & \hatcurXAv{}                \\            
~~~~Distance (pc)\dotfill                & \hatcurXdistred{}\phn          & \hatcurXdistredhpsmodel{}   \\ 
\sidehead{Binary Star Companion \hatcur{}B}
~~~~$\mstar$ ($\msun$)\dotfill           & $\cdots$                       & \hatcurISOmlongBhpsmodel{}  \\
~~~~$\rstar$ ($\rsun$)\dotfill           & $\cdots$                       & \hatcurISOrlongBhpsmodel{}  \\
~~~~$\loggstar$ (cgs)\dotfill            & $\cdots$                       & \hatcurISOloggBhpsmodel{}   \\
~~~~$\lstar$ ($\lsun$)\dotfill           & $\cdots$                       & \hatcurISOlumBhpsmodel{}    \\
~~~~$\teffstar$ (K)\dotfill              & $\cdots$                       & \hatcurISOteffBhpsmodel{}   \\  
\enddata
\tablecomments{
	The listed parameters for the ``Single Star'' model are those
	determined through the joint
	differential evolution Markov Chain analysis described in
	Section~\ref{sec:globmod}, while the ``Binary Star'' model paramters
	are determined as described in Section~\ref{sec:blend}. Systematic
	errors in the bolometric correction tables or stellar evolution
	models are not included, and likely dominate the error budget. The
	``Single Star'' values are determined based on an empirical method,
	while the ``Binary Star'' values make use of stellar evolution
	models. In the latter case, the constraint from these models leads
	to very low formal uncertainties on parameters such as \teffstar.
}
\ifthenelse{\boolean{emulateapj}}{
    \end{deluxetable*}
}{
    \end{deluxetable}
}
\clearpage

\startlongtable
\ifthenelse{\boolean{emulateapj}}{
    \begin{deluxetable*}{lcc}
}{
    \begin{deluxetable}{lcc}
}
\tabletypesize{\tiny}
\tablecaption{Orbital and planetary parameters for \hatcurb{}\label{tab:planetparam}}
\tablehead{
    \multicolumn{1}{c}{~~~~~~~~~~~~~~~Parameter~~~~~~~~~~~~~~~} &
    \multicolumn{1}{c}{Value} &
    \multicolumn{1}{c}{Value} \\
    \multicolumn{1}{c}{} &
    \multicolumn{1}{c}{Single Star} &
    \multicolumn{1}{c}{Binary Star}
}
\startdata
\noalign{\vskip -3pt}
\sidehead{\Lc{} parameters}
~~~$P$ (days)             \dotfill                           & $\hatcurLCP{}$                 & $\hatcurLCPhpsmodel{}$     \\
~~~$T_c$ (${\rm BJD}$)\tablenotemark{a}   \dotfill           & $\hatcurLCT{}$                 & $\hatcurLCThpsmodel{}$     \\
~~~$T_{14}$ (days)\tablenotemark{a}   \dotfill               & $\hatcurLCdur{}$               & $\hatcurLCdurhpsmodel{}$   \\
~~~$T_{12} = T_{34}$ (days)\tablenotemark{a} \dotfill        & $\hatcurLCingdur{}$            & $\hatcurLCingdurhpsmodel{}$\\
~~~$\arstar$              \dotfill                           & $\hatcurPPar{}$                & $\hatcurPParhpsmodel{}$    \\
~~~$\zrstar$ \tablenotemark{b}             \dotfill          & $\hatcurLCzeta{}$\phn          & $\hatcurLCzetahpsmodel{}$\phn \\
~~~$\rpl/\rstar$          \dotfill                           & $\hatcurLCrprstar{}$           & $\hatcurLCrprstarhpsmodel{}$\\
~~~$b^2$                  \dotfill                           & $\hatcurLCbsq{}$               & $\hatcurLCbsqhpsmodel{}$   \\
~~~$b \equiv a \cos i/\rstar$ \dotfill                       & $\hatcurLCimp{}$               & $\hatcurLCimphpsmodel{}$   \\
~~~$i$ (deg)              \dotfill                           & $\hatcurPPi{}$\phn             & $\hatcurPPihpsmodel{}$\phn \\
%
%
\sidehead{Limb-darkening coefficients \tablenotemark{c}}
~~~$c_1,r$                  \dotfill    & $\hatcurLBir{}$ & $\hatcurLBirhpsmodel{}$   \\
~~~$c_2,r$                  \dotfill    & $\hatcurLBiir{}$ & $\hatcurLBiirhpsmodel{}$ \\
~~~$c_1,R$                  \dotfill    & $\hatcurLBiR{}$ & $\hatcurLBiRhpsmodel{}$   \\
~~~$c_2,R$                  \dotfill    & $\hatcurLBiiR{}$ & $\hatcurLBiiRhpsmodel{}$ \\
~~~$c_1,i$                  \dotfill    & $\hatcurLBii{}$ & $\hatcurLBiihpsmodel{}$   \\
~~~$c_2,i$                  \dotfill    & $\hatcurLBiii{}$ & $\hatcurLBiiihpsmodel{}$ \\
~~~$c_1,I$                  \dotfill    & $\hatcurLBiI{}$ & $\hatcurLBiIhpsmodel{}$   \\
~~~$c_2,I$                  \dotfill    & $\hatcurLBiiI{}$ & $\hatcurLBiiIhpsmodel{}$ \\
\sidehead{RV parameters \tablenotemark{d}}
~~~$K$ (\ms)              \dotfill    & $\hatcurRVK{}$\phn\phn & $\hatcurRVKhpsmodel{}$ \\
%
%
%
~~~RV jitter PFS (\ms) \tablenotemark{e}       \dotfill    & $\hatcurRVjitter{}$ & $\cdots$ \\
\sidehead{Planetary parameters}
~~~$\mpl$ ($\mjup$)       \dotfill                           & $\hatcurPPmlong{}$             & $\hatcurPPmlonghpsmodel$ \\   
~~~$\rpl$ ($\rjup$)       \dotfill                           & $\hatcurPPrlong{}$             & $\hatcurPPrlonghpsmodel{}$ \\ 
~~~$C(\mpl,\rpl)$\tablenotemark{f}\dotfill                   & $\hatcurPPmrcorr{}$            & $\cdots$ \\                   
~~~$\rhopl$ (\gcmc)       \dotfill                           & $\hatcurPPrho{}$               & $\hatcurPPrhohpsmodel{}$ \\   
~~~$\log g_p$ (cgs)       \dotfill                           & $\hatcurPPlogg{}$              & $\hatcurPPlogghpsmodel{}$ \\  
~~~$a$ (AU)               \dotfill                           & $\hatcurPParel{}$              & $\hatcurPParelhpsmodel$ \\    
~~~$T_{\rm eq}$ (K)        \dotfill                          & $\hatcurPPteff{}$              & $\hatcurPPteffhpsmodel{}$ \\  
~~~$\Theta$ \tablenotemark{g}\dotfill                        & $\hatcurPPtheta{}$             & $\hatcurPPthetahpsmodel{}$ \\ 
%
~~~$\log_{10}\langle F \rangle$ (cgs) \tablenotemark{h} \dotfill    & $\hatcurPPfluxavglog{}$ & $\hatcurPPfluxavgloghpsmodel{}$ \\
\enddata
\tablecomments{
	Parameters in the ``Single Star'' column are determined as
	described in \refsecl{globmod} assuming \hatcur{} is a single star
	with a transiting planet.  Parameters listed in the ``Binary Star''
	column are determined as described in \refsecl{blend} assuming
	\hatcur{} is an unresolved binary star with a transiting planet
	around one component.  In both cases we assume the orbit is
	circular.
}
\tablenotetext{a}{
    Times are in Barycentric Julian Date computed on the TDB system
	with correction for leap seconds.
    \ensuremath{T_c}: Reference epoch of mid transit that minimizes the
    correlation with the orbital period.
    \ensuremath{T_{12}}: total transit duration, time between first to
    last contact;
    \ensuremath{T_{12}=T_{34}}: ingress/egress time, time between first
    and second, or third and fourth contact.
}
\tablenotetext{b}{
   Reciprocal of the half duration of the transit used as a jump
   parameter in our MCMC analysis in place of $\arstar$.  It is related
   to $\arstar$ by the expression $\zrstar =
   \arstar(2\pi(1+e\sin\omega))/(P\sqrt{1-b^2}\sqrt{1-e^2})$
   \citep{bakos:2010:hat11}.
}
\tablenotetext{c}{
    Values for a quadratic law, adopted from the tabulations by
    \cite{claret:2004}.  For the ``Single Star'' model these are fixed
    according to the spectroscopic parameters determined from the
    ARCoIRIS spectrum.  For the ``Binary Star'' model these are varied
    at each step in the Markov Chain as the atmospheric parameters of
    the model star are varied, here we list the median parameter
    values.  For the {\em TESS} light curves we assume the $I$-band
    limb darkening coefficients on the grounds that the unfiltered TESS
    bandpass is dominated by light from that portion of the spectrum
    for this M dwarf.
}
\tablenotetext{d}{
    The ``Binary Star'' model value for $K$ is based on the ``Single
    Star'' model value scaled by a factor of $1.16 \pm 0.23$ to account
    for dilution from the binary star companion as described in
    Section~\ref{sec:blend}.
}
\tablenotetext{e}{
    Term added in quadrature to the formal RV uncertainties for each
    instrument. This is treated as a free parameter in the fitting
    routine for the ``Single Star'' model. 
}
\tablenotetext{e}{
    Correlation coefficient between the planetary mass \mpl\ and radius
    \rpl\ estimated from the posterior parameter distribution.  This
    was not estimated for the ``Binary Star'' model.
}
\tablenotetext{f}{
    The Safronov number is given by $\Theta = \frac{1}{2}(V_{\rm
    esc}/V_{\rm orb})^2 = (a/\rpl)(\mpl / \mstar )$
    \citep[see][]{hansen:2007}.
}
\tablenotetext{g}{
    Incoming flux per unit surface area, averaged over the orbit.
}
\ifthenelse{\boolean{emulateapj}}{
    \end{deluxetable*}
}{
    \end{deluxetable}
}

\clearpage

\tabletypesize{\scriptsize}
\ifthenelse{\boolean{emulateapj}}{
    \begin{deluxetable*}{rrrrrr}
}{
    \begin{deluxetable}{rrrrrr}
}
\tablewidth{0pc}
\tablecaption{
    Relative radial velocities and bisector spans from PFS/Magellan for
    \hatcur{}.  \label{tab:rvs}
}
\tablehead{
    \colhead{BJD} &
    \colhead{RV\tablenotemark{a}} &
    \colhead{\ensuremath{\sigma_{\rm RV}}\tablenotemark{b}} &
    \colhead{BS} &
    \colhead{\ensuremath{\sigma_{\rm BS}}} &
    \colhead{Phase}\\
    \colhead{\hbox{(2,450,000$+$)}} &
    \colhead{(\ms)} &
    \colhead{(\ms)} &
    \colhead{(\ms)} &
    \colhead{(\ms)} &
    \colhead{}
}
\startdata
\input{rvtable.tex}
\enddata
\tablenotetext{a}{
    The zero-point of these velocities is arbitrary. An overall offset
    $\gamma$ fitted to the velocities has been subtracted.
}
\tablenotetext{b}{
    Internal errors excluding the component of astrophysical jitter
    considered in \refsecl{globmod}.
}
\ifthenelse{\boolean{emulateapj}}{
    \end{deluxetable*}
}{
    \end{deluxetable}
}

\end{document}

%% file: newcommand_gen.tex
\newcommand{\forloop}[5][1]%
{%
\setcounter{#2}{#3}%
\ifthenelse{#4}%
	{%
	#5%
	\addtocounter{#2}{#1}%
	\forloop[#1]{#2}{\value{#2}}{#4}{#5}%
	}%
	{%
	}%
}%

\newcommand{\ctbd}[1]{}

\newcommand{\lcs}{light curves}
\newcommand{\Lc}{Light curve}

\newcommand{\masy}{\ensuremath{\rm mas\,yr^{-1}}}
\newcommand{\kms}{\ensuremath{\rm km\,s^{-1}}}
\newcommand{\ms}{\ensuremath{\rm m\,s^{-1}}}

\newcommand{\gcmc}{\ensuremath{\rm g\,cm^{-3}}}

\newcommand{\logg}{\ensuremath{\log{g}}}
\newcommand{\vsini}{\ensuremath{v \sin{i}}}
\newcommand{\feh}{\ensuremath{\rm [Fe/H]}}

\newcommand{\rsun}{\ensuremath{R_\sun}}
\newcommand{\msun}{\ensuremath{M_\sun}}
\newcommand{\lsun}{\ensuremath{L_\sun}}

\newcommand{\rstar}{\ensuremath{R_\star}}
\newcommand{\mstar}{\ensuremath{M_\star}}
\newcommand{\lstar}{\ensuremath{L_\star}}

\newcommand{\teffstar}{\ensuremath{T_{\rm eff\star}}}
\newcommand{\rhostar}{\ensuremath{\rho_\star}}
\newcommand{\loggstar}{\ensuremath{\log{g_{\star}}}}

\newcommand{\rpl}{\ensuremath{R_{p}}}
\newcommand{\mpl}{\ensuremath{M_{p}}}

\newcommand{\rhopl}{\ensuremath{\rho_{p}}}

\newcommand{\arstar}{\ensuremath{a/\rstar}}
\newcommand{\zrstar}{\ensuremath{\zeta/\rstar}}

\newcommand{\rjup}{\ensuremath{R_{\rm J}}}
\newcommand{\mjup}{\ensuremath{M_{\rm J}}}

\newcommand{\reffig}[1]{Fig.~\ref{fig:#1}}

\newcommand{\reffigl}[1]{Figure~\ref{fig:#1}}
\newcommand{\refsecl}[1]{\mbox{Section \ref{sec:#1}}}

\newcommand{\reftabl}[1]{Table~\ref{tab:#1}}



%% file: HATS755-002.tex
\newcommand{\hatcurhtr}{HATS755-002}                                   
\newcommand{\hatcurCCra}{\ensuremath{01^{\mathrm h}02^{\mathrm m}12.2812{\mathrm s}}}                                
\newcommand{\hatcurCCdec}{\ensuremath{-61{\arcdeg}45{\arcmin}21.6599{\arcsec}}}                              
\newcommand{\hatcurCCgsc}{GSC~}                                        
\newcommand{\hatcurCCtassmg}{\ensuremath{17.105\pm0.042}}              
\newcommand{\hatcurCCtassmr}{\ensuremath{15.8100\pm0.0090}}            
\newcommand{\hatcurCCtassmi}{\ensuremath{14.575\pm0.031}}              
\newcommand{\hatcurCCparallax}{\ensuremath{7.103\pm0.043}}             
\newcommand{\hatcurCCgaiamG}{\ensuremath{15.35120\pm0.00050}}          
\newcommand{\hatcurCCgaiamBP}{\ensuremath{16.7435\pm0.0038}}           
\newcommand{\hatcurCCgaiamRP}{\ensuremath{14.1873\pm0.0013}}           
\newcommand{\hatcurCCtwomassJmag}{\ensuremath{12.605\pm0.026}}         
\newcommand{\hatcurCCtwomassHmag}{\ensuremath{11.972\pm0.032}}         
\newcommand{\hatcurCCtwomassKmag}{\ensuremath{11.727\pm0.025}}         
\newcommand{\hatcurLCdip}{\ensuremath{61.8}}                           
\newcommand{\hatcurLCrprstar}{\ensuremath{0.2155\pm0.0017}} 
\newcommand{\hatcurLCbsq}{\ensuremath{0.108_{-0.034}^{+0.035}}} 
\newcommand{\hatcurLCimp}{\ensuremath{0.329_{-0.056}^{+0.050}}} 
\newcommand{\hatcurLCzeta}{\ensuremath{28.74\pm0.18}}     
\newcommand{\hatcurLCdur}{\ensuremath{0.08622\pm0.00077}} 
\newcommand{\hatcurLCingdur}{\ensuremath{0.01686\pm0.00067}} 
\newcommand{\hatcurLCP}{\ensuremath{3.7955213\pm0.0000011}} 
\newcommand{\hatcurLCPshort}{\ensuremath{3.7955}}         
\newcommand{\hatcurLCT}{\ensuremath{2457699.38939\pm0.00020}} 
\newcommand{\hatcurLCrho}{\ensuremath{5.80\pm0.31}}       
\newcommand{\hatcurSMEiteff}{\ensuremath{3500\pm120}}                  
\newcommand{\hatcurSMEizfeh}{\ensuremath{0.26\pm0.13}}                 
\newcommand{\hatcurSMEizfehshort}{\ensuremath{0.26}}                   
\newcommand{\hatcurSMEilogg}{\ensuremath{4.50\pm0.50}}                 
\newcommand{\hatcurSMEivsin}{\ensuremath{nff\pmnff}}                   
\newcommand{\hatcurSMEivmac}{\ensuremath{nff\pmnff}}                   
\newcommand{\hatcurSMEivmic}{\ensuremath{nff\pmnff}}                   
\newcommand{\hatcurLBii}{\ensuremath{0.3460}}                          
\newcommand{\hatcurLBiii}{\ensuremath{0.3337}}                         
\newcommand{\hatcurLBiI}{\ensuremath{0.3128}}                          
\newcommand{\hatcurLBiiI}{\ensuremath{0.3638}}                         
\newcommand{\hatcurLBir}{\ensuremath{0.5401}}                          
\newcommand{\hatcurLBiir}{\ensuremath{0.2354}}                         
\newcommand{\hatcurLBiR}{\ensuremath{0.4889}}                          
\newcommand{\hatcurLBiiR}{\ensuremath{0.2566}}                         
\newcommand{\hatcurISOmlong}{\ensuremath{0.569_{-0.069}^{+0.042}}} 
\newcommand{\hatcurISOrlong}{\ensuremath{0.5161_{-0.0099}^{+0.0053}}} 
\newcommand{\hatcurISOlogg}{\ensuremath{4.766\pm0.027}}   
\newcommand{\hatcurISOlum}{\ensuremath{0.02702\pm0.00035}} 
\newcommand{\hatcurISOfeh}{\ensuremath{0.237\pm0.063}}    
\newcommand{\hatcurISOteff}{\ensuremath{3254\pm22}}       
\newcommand{\hatcurRVK}{\ensuremath{87\pm44}}             
\newcommand{\hatcurRVjitter}{\ensuremath{91\pm42}}        
\newcommand{\hatcurPPi}{\ensuremath{88.85\pm0.18}}        
\newcommand{\hatcurPPlogg}{\ensuremath{2.98_{-0.27}^{+0.19}}} 
\newcommand{\hatcurPPar}{\ensuremath{16.41\pm0.29}}       
\newcommand{\hatcurPParel}{\ensuremath{0.03946_{-0.00168}^{+0.00095}}} 
\newcommand{\hatcurPPrho}{\ensuremath{0.44\pm0.23}}       
\newcommand{\hatcurPPmlong}{\ensuremath{0.45\pm0.24}}     
\newcommand{\hatcurPPrlong}{\ensuremath{1.080\pm0.016}}   
\newcommand{\hatcurPPmrcorr}{\ensuremath{0.05}}                        
\newcommand{\hatcurPPteff}{\ensuremath{567.9_{-6.6}^{+12.6}}} 
\newcommand{\hatcurPPtheta}{\ensuremath{0.059\pm0.030}}   
\newcommand{\hatcurPPfluxavglog}{\ensuremath{7.373_{-0.020}^{+0.038}}} 
\newcommand{\hatcurXAv}{\ensuremath{0.032\pm0.011}}       
\newcommand{\hatcurXdistred}{\ensuremath{140.68\pm0.83}}  
\newcommand{\hatcurCCpmra}{\ensuremath{78.858\pm0.087}}                
\newcommand{\hatcurCCpmdec}{\ensuremath{-27.095\pm0.064}}              

%% file: HATS755-002.hps.tex
\newcommand{\hatcurLCrprstarhpsmodel}{\ensuremath{0.2297\pm0.0031}}    
\newcommand{\hatcurLCbsqhpsmodel}{\ensuremath{0.013_{-0.011}^{+0.015}}} 
\newcommand{\hatcurLCimphpsmodel}{\ensuremath{0.112_{-0.068}^{+0.053}}} 
\newcommand{\hatcurLCzetahpsmodel}{\ensuremath{28.63\pm0.17}}          
\newcommand{\hatcurLCdurhpsmodel}{\ensuremath{0.08612\pm0.00044}}      
\newcommand{\hatcurLCingdurhpsmodel}{\ensuremath{0.01629\pm0.00020}}   
\newcommand{\hatcurLCPhpsmodel}{\ensuremath{3.7955203\pm0.0000012}}    
\newcommand{\hatcurLCThpsmodel}{\ensuremath{2457699.38955\pm0.00024}}  
\newcommand{\hatcurLCrhohpsmodel}{\ensuremath{6.65\pm0.17}}            
\newcommand{\hatcurISOmlonghpsmodel}{\ensuremath{0.4651\pm0.0062}}     
\newcommand{\hatcurISOrlonghpsmodel}{\ensuremath{0.4618\pm0.0057}}     
\newcommand{\hatcurISOlogghpsmodel}{\ensuremath{4.7763\pm0.0063}}      
\newcommand{\hatcurISOlumhpsmodel}{\ensuremath{0.02249\pm0.00053}}     
\newcommand{\hatcurISOfehhpsmodel}{\ensuremath{0.305\pm0.036}}         
\newcommand{\hatcurISOteffhpsmodel}{\ensuremath{3294.3\pm4.3}}         
\newcommand{\hatcurISOagehpsmodel}{\ensuremath{7.7_{-4.8}^{+3.5}}}     
\newcommand{\hatcurISOmlongBhpsmodel}{\ensuremath{0.243\pm0.013}}      
\newcommand{\hatcurISOrlongBhpsmodel}{\ensuremath{0.2714\pm0.0099}}    
\newcommand{\hatcurISOloggBhpsmodel}{\ensuremath{4.9575\pm0.0097}}     
\newcommand{\hatcurISOlumBhpsmodel}{\ensuremath{0.00473\pm0.00058}}    
\newcommand{\hatcurISOteffBhpsmodel}{\ensuremath{2912\pm37}}           
\newcommand{\hatcurPPihpsmodel}{\ensuremath{89.63\pm0.18}}             
\newcommand{\hatcurPParhpsmodel}{\ensuremath{17.17\pm0.15}}            
\newcommand{\hatcurPParelhpsmodel}{\ensuremath{0.03689\pm0.00016}}     
\newcommand{\hatcurPPrlonghpsmodel}{\ensuremath{1.032\pm0.010}}        
\newcommand{\hatcurPPteffhpsmodel}{\ensuremath{562.0\pm2.3}}           
\newcommand{\hatcurPPfluxavgloghpsmodel}{\ensuremath{7.3520\pm0.0071}} 
\newcommand{\hatcurXdistredhpsmodel}{\ensuremath{140.75\pm0.86}}       
\newcommand{\hatcurRVKhpsmodel}{\ensuremath{98\pm56}}                  
\newcommand{\hatcurPPlogghpsmodel}{\ensuremath{3.02_{-0.29}^{+0.21}}}  
\newcommand{\hatcurPPrhohpsmodel}{\ensuremath{0.51\pm0.29}}            
\newcommand{\hatcurPPmtwosiglimhpsmodel}{\ensuremath{<0.949}}          
\newcommand{\hatcurPPmlonghpsmodel}{\ensuremath{0.45\pm0.26}}          
\newcommand{\hatcurPPthetahpsmodel}{\ensuremath{0.069\pm0.039}}        

%% file: phfu_tab_short.tex
$ 56194.46534 $ & $  14.46785 $ & $   0.02515 $ & $  -0.05133 $ & $ r$ &  HS/G755.4\\
$ 56183.07906 $ & $  14.53916 $ & $   0.02462 $ & $   0.01998 $ & $ r$ &  HS/G755.4\\
$ 56202.05668 $ & $  14.48734 $ & $   0.03641 $ & $  -0.03184 $ & $ r$ &  HS/G755.4\\
$ 56167.89733 $ & $  14.49451 $ & $   0.02565 $ & $  -0.02467 $ & $ r$ &  HS/G755.4\\
$ 56141.32989 $ & $  14.51297 $ & $   0.03681 $ & $  -0.00621 $ & $ r$ &  HS/G755.4\\
$ 56213.44497 $ & $  14.51480 $ & $   0.02061 $ & $  -0.00438 $ & $ r$ &  HS/G755.4\\
$ 56114.76192 $ & $  14.58036 $ & $   0.04107 $ & $   0.06118 $ & $ r$ &  HS/G755.4\\
$ 56186.87727 $ & $  14.53484 $ & $   0.02796 $ & $   0.01566 $ & $ r$ &  HS/G755.4\\
$ 56145.12690 $ & $  14.46642 $ & $   0.02888 $ & $  -0.05276 $ & $ r$ &  HS/G755.4\\
$ 56167.90064 $ & $  14.51266 $ & $   0.02463 $ & $  -0.00652 $ & $ r$ &  HS/G755.4\\

%% file: rvtable.tex
$ 7022.57729 $ & $    41.90 $ & $    16.56 $ & $ -264.3 $ & $ 1123.8 $ & $   0.681 $ \\
$ 7025.56597 $ & $   -46.25 $ & $    15.91 $ & $ -816.3 $ & $ 1469.2 $ & $   0.469 $ \\
$ 7026.58472 $ & \nodata      & \nodata      & $   33.9 $ & $  932.1 $ & $   0.737 $ \\
$ 7325.69973 $ & $  -105.95 $ & $    19.13 $ & \nodata      & \nodata      & $   0.545 $ \\
$ 7385.55818 $ & $   -69.45 $ & $    17.33 $ & $    6.7 $ & $  970.8 $ & $   0.315 $ \\
$ 7614.83615 $ & $   189.26 $ & $    17.52 $ & $ -220.5 $ & $  358.3 $ & $   0.723 $ \\
$ 7616.85346 $ & $    19.70 $ & $    15.83 $ & $ 2859.9 $ & $  969.4 $ & $   0.254 $ \\
$ 7766.54828 $ & $   120.67 $ & $    22.78 $ & $ 3214.5 $ & $ 2049.4 $ & $   0.694 $ \\